\documentclass[aps, prd, twocolumn, nofootinbib,]{revtex4-2}

\usepackage{amsmath}
\usepackage{amsfonts}
\usepackage{wasysym}
\usepackage{graphicx}
\usepackage{comment}
\usepackage{tensor}
\usepackage{hyphenat}

\def\filetype{pdf}

\def\path{}

\allowdisplaybreaks[1]

\begin{document}


\title{Fermion--charged-boson stars}
\author{Ben Kain}
\affiliation{Department of Physics, College of the Holy Cross, Worcester, Massachusetts 01610, USA}

\begin{abstract}
\noindent Fermion-boson stars are starlike systems composed of the ordinary nuclear matter of a neutron star and bosonic dark matter.  The bosonic dark matter has typically been taken to be a complex scalar field.  A natural extension is for the complex scalar field to be charged.  We make the simplest extension and gauge the scalar field under $U(1)$.  We therefore study fermion--charged-boson stars.  We make a detailed study of the stability of this system by computing critical curves over the whole of parameter space.  We then study how the fermion and boson sectors contribute to the total mass of the star.  Finally, we present mass\hyp radius diagrams, showing that an increase in charge can lead to more massive and more compact stars.
\end{abstract} 

\maketitle


\section{Introduction}

Fermion-boson stars are starlike configurations made up of a mixture of fermions and bosons \cite{Henriques:1989ar}.  The fermion sector is described by a perfect fluid energy\hyp momentum tensor and an equation of state \cite{GlendenningBook}.  Typically, one uses an equation of state that can describe the nuclear matter in a neutron star. The boson sector is most commonly taken to be a complex scalar field.  By itself, a complex scalar field can form boson stars \cite{Kaup:1968zz, Ruffini:1969qy, Jetzer:1991jr, Schunck:2003kk, Liebling:2012fv}.

Dark matter direct detection experiments have placed stringent constraints on the dark matter-nucleon coupling strength \cite{Akerib:2017kat, Cui:2017nnn, Aprile:2018dbl}.  From the perspective of fermion-boson stars, it is generally assumed that the coupling strength is negligibly small and that there exists only gravitational interactions between the fermion and boson sectors.

Fermion-boson stars are therefore examples of dark matter admixed neutron stars \cite{Henriques:1989ar, Sandin:2008db, Leung:2011zz} with bosonic dark matter.  It is a theoretical possibility that during the formation process of a neutron star, dark matter could become mixed with ordinary matter in sufficient quantities to affect bulk properties of the star, such as its mass and radius.  Neutron stars, then, offer the exciting possibility of indirectly probing dark matter.

Fermion-boson stars were first studied by Henriques, Liddle, and Moorhouse \cite{Henriques:1989ar, Henriques:1989ez, Henriques:1990xg}.  In these works, the fermion sector was a free Fermi gas of neutrons and the boson sector was a neutral complex scalar field.  More recently, these systems have been evolved with full numerical relativity, with the fermion sector described by a polytropic equation of state and the boson sector again taken to be a neutral complex scalar field \cite{ValdezAlvarado:2012xc, DiGiovanni:2020frc, Valdez-Alvarado:2020vqa, DiGiovanni:2021vlu}.  For additional work on fermion-boson stars, see \cite{deSousa:1995ye, Pisano:1995yk, Sakamoto:1998aj, deSousa:2000eq, Henriques:2003yr, Dzhunushaliev:2011ma, Brito:2015yga, Brito:2015yfh}.

Since the scalar field is complex, a natural extension is for the scalar field to be charged.  We make the simplest extension and gauge the scalar field under $U(1)$ so that the scalar field interacts with dark photons.  By themselves, such charged scalar fields can form charged boson stars \cite{Jetzer:1989av, Pugliese:2013gsa}.  When mixed with fermions, then, they can form fermion--charged-boson stars.

A useful dimensionless quantity for understanding the relative mass scales for the fermion and boson sectors can be shown to be \cite{Henriques:1989ar}
\begin{equation}
\beta \equiv \frac{(\mu_N/m_P)^2}{\mu/m_P},
\end{equation}
where $m_P$ is the Planck mass, $\mu$ is the scalar field particle mass, and $\mu_N$ is the representative particle mass for the fermion sector, which we take to be the nucleon mass $\mu_N = 938$ MeV.  Roughly speaking, when $\beta$ is order one, there is a single mass scale in the system and the fermion and boson sectors can both be non\hyp negligible \cite{Henriques:1989ar}.  For order one $\beta$, we have $\mu \sim 10^{-10}$ eV.  Thus, fermion-boson stars are most interesting with ultralight bosonic dark matter \cite{Henriques:1989ar, Sin:1992bg, Chavanis:2011cz, Hu:2000ke}.  This is not unexpected, since it is well\hyp known that boson stars take on astrophysical sizes, with masses in the solar mass range and radii in the kilometer range, when the scalar field mass is ultralight \cite{Liebling:2012fv}.

In this paper, we study static solutions of spherically symmetric fermion--charged-boson stars.  We use a realistic equation of state in the fermion sector.  In the boson sector, we focus on a scalar field with mass $10^{-10}$ eV, but at times will also consider larger masses.  In Sec.\ \ref{sec:metric}, we present two parametrizations for the spherically symmetric metric that we will be using and the equations for determining the metric functions.  As mentioned above, we assume that there exists only gravitational interactions between the fermion and boson sectors.  Consequently, the equations of motion for the two sectors can be derived independently.  In Sec. \ref{sec:FS}, we review fermion stars, for which the equations of motion are the Tolman\hyp Oppenheimer\hyp Volkoff (TOV) equations.  In doing this, we use the NL3 equation of state \cite{Lalazissis:1996rd, ToddRutel:2005zz}, which is a realistic equation of state for the nuclear matter in a neutron star.  We detail how this equation of state is constructed in the Appendix.  In Sec.\ \ref{sec:CBS}, we review charged boson stars.  We take special care to explain how a particular gauge choice leads to all fields being time\hyp independent.  In Sec.\ \ref{sec:FCBS}, we combine the equations presented in Secs.\ \ref{sec:FS} and \ref{sec:CBS} and study fermion--charged-boson stars.  In addition to solving the equations for static solutions and presenting mass-radius diagrams, there are two important questions we address.  The first concerns the stability of the static solutions.  We use the two\hyp fluid methods developed in \cite{Henriques:1990xg} and determine stability over the whole of parameter space.  The second question concerns the relative contributions from the fermion and boson sectors.  We study this issue by looking at the relative contributions to the total mass of the star.  In this way, we show that as the mass of the scalar field particle is increased, the transition from being fermion dominated to being boson dominated sharpens.  We conclude in Sec.\ \ref{sec:conclusion}.


\section{Metric and metric equations}
\label{sec:metric}

We focus on spherically symmetric spacetimes and use units such that $c=\hbar =1$.  We will make use of two parametrizations of the spherically symmetric metric.  One parametrization is
\begin{equation}
ds^2 = -\alpha^2 dt^2 + a^2 dr^2 + r^2(d\theta^2 +\sin^2\theta d\phi^2),
\end{equation}
where $\alpha$ and $a$ are metric functions determined from the Einstein field equations,
\begin{equation}
G\indices{^\mu_\nu} = 8\pi G (T_\text{tot})\indices{^\mu_\nu},
\end{equation}
where $G\indices{^\mu_\nu}$ is the Einstein tensor, $G$ is the gravitational constant, and $(T_\text{tot})\indices{^\mu_\nu}$ is the energy-momentum tensor.  

In Sec.\ \ref{sec:FCBS}, we consider fermion--charged-boson stars and $(T_\text{tot})\indices{^\mu_\nu}$ will have both fermionic and bosonic contributions,
\begin{equation} \label{T tot}
(T_\text{tot})\indices{^\mu_\nu} = (T_f)\indices{^\mu_\nu} + (T_b)\indices{^\mu_\nu}.
\end{equation}
That the two contributions separate follows from our assumption that there exists only gravitational interactions between the fermion and boson sectors.  In Sec.\ \ref{sec:FS}, we review fermion stars, in which the energy\hyp momentum tensor has only the fermionic contribution, and in Sec.\ \ref{sec:CBS}, we review charged bosons stars, in which the energy\hyp momentum tensor has only the bosonic contribution.  Since the two contributions in Eq.\ (\ref{T tot}) are independently conserved, $\nabla_\mu (T_f)\indices{^\mu_\nu} = 0$ and $\nabla_\mu (T_b)\indices{^\mu_\nu} = 0$, where $\nabla_\mu$ is the metric-covariant derivative, the equations derived in the next two sections will be directly applicable to fermion--charged-boson stars.

The Einstein field equations lead to the following equations for determining $\alpha$ and $a$,
\begin{equation} \label{metric equations}
\begin{split}
\frac{\alpha'}{\alpha} &= +4\pi G r  a^2 (T_\text{tot})\indices{^r_r} + \frac{a^2-1}{2r}
\\
\frac{a'}{a} &= 
-4\pi G r a^2 (T_\text{tot})\indices{^t_t}
- \frac{a^2 - 1}{2r} 
\\
\frac{\dot{a}}{a} &= -4\pi G r \alpha^2 (T_\text{tot})\indices{^t_r},
\end{split}
\end{equation}
where a prime denotes an $r$ derivative and a dot denotes a $t$ derivative.

When solving for static solutions, an alternative parametrization of the metric is useful, which is
\begin{equation}
ds^2 = -\sigma^2 N dt^2 + \frac{dr^2}{N} + r^2(d\theta^2 +\sin^2\theta d\phi^2),
\end{equation}
where $N \equiv 1 - 2 G m/r$ and the new metric functions are
\begin{equation} \label{sigma m def}
\sigma \equiv \alpha a, \qquad
m \equiv \frac{r}{2G} \left(1 - \frac{1}{a^2} \right).
\end{equation}
Using the first two equations in (\ref{metric equations}), $\sigma$ and $m$ obey
\begin{equation} \label{sigma m eqs}
\begin{split}
\frac{\sigma'}{\sigma} &= \frac{4\pi G r}{N} \left[ (T_\text{tot})\indices{^r_r} - (T_\text{tot})\indices{^t_t} \right]
\\
m' &= -4\pi r^2 (T_\text{tot})\indices{^t_t}.
\end{split}
\end{equation}
We will also need
\begin{equation} \label{alpha prime eq}
\frac{\alpha'}{\alpha} = \frac{4\pi G r^3(T_\text{tot})\indices{^r_r} + G m}{r^2 N},
\end{equation}
which follows directly from the first equation in (\ref{metric equations}).


\section{Fermion stars}
\label{sec:FS}

In this section, we review fermion stars.  The equations derived here will be directly applicable to fermion--charged-boson stars.  Our interest is with static solutions, for which spacetime is time\hyp independent.  For fermion stars, this immediately leads to all quantities being time\hyp independent (things are not as straightforward for charged bosons stars, as we will see in the next section).  

Fermion stars are described by a perfect fluid energy\hyp momentum tensor \cite{GlendenningBook},
\begin{equation} \label{Tf}
(T_f)\indices{^\mu_\nu} = \text{diag} (-\epsilon, p, p, p),
\end{equation}
where $\epsilon$ is the energy density and $p$ is the pressure of the fluid.  Conservation of the energy\hyp momentum tensor, $\nabla_\mu (T_f)\indices{^\mu_\nu} = 0$, leads to the equation of motion,
\begin{equation} \label{f eom}
p' = - \frac{\alpha'}{\alpha} (\epsilon + p).
\end{equation}
Equation (\ref{f eom}) and Eqs.\ (\ref{sigma m eqs}) and (\ref{alpha prime eq}) with $(T_\text{tot})\indices{^\mu_\nu} = (T_f)\indices{^\mu_\nu}$ are the TOV equations.  Note that $\sigma$ decouples and does not have to be computed if it is not needed.

In addition to the energy\hyp momentum tensor, fermion stars are described by an equation of state, $\epsilon = \epsilon(p)$. We use the NL3 equation of state \cite{Lalazissis:1996rd, ToddRutel:2005zz}, which is a realistic equation of state for neutron stars.  We do not have a particular reason for choosing NL3, aside from it being realistic, and we expect the qualitative aspects of our results to hold for other choices.  Nevertheless, for completeness, we review the construction of NL3 in the Appendix.

To solve the TOV equations, we need inner boundary conditions.  Inner boundary conditions can be obtained by plugging Taylor expansions of the fields into the TOV equations, which gives $p(r) = p(0) + O(r^2)$ and $m(r)=O(r^3)$.  Fermion stars are uniquely identified by the central pressure, $p(0)$.  We thus specify a central pressure and then integrate  Eq.\ (\ref{f eom}) and the bottom equation in (\ref{sigma m eqs}) outward from $r = 0$.  At each integration step we know the pressure and we compute the energy density using the equation of state.  The radius of the star, $R_f$, is defined by the radial position at which the pressure hits zero, $p(R_f) = 0$, and the mass of the star is given by $M = m(R_f)$.

\begin{figure}
\centering
\includegraphics[width=3.4in]{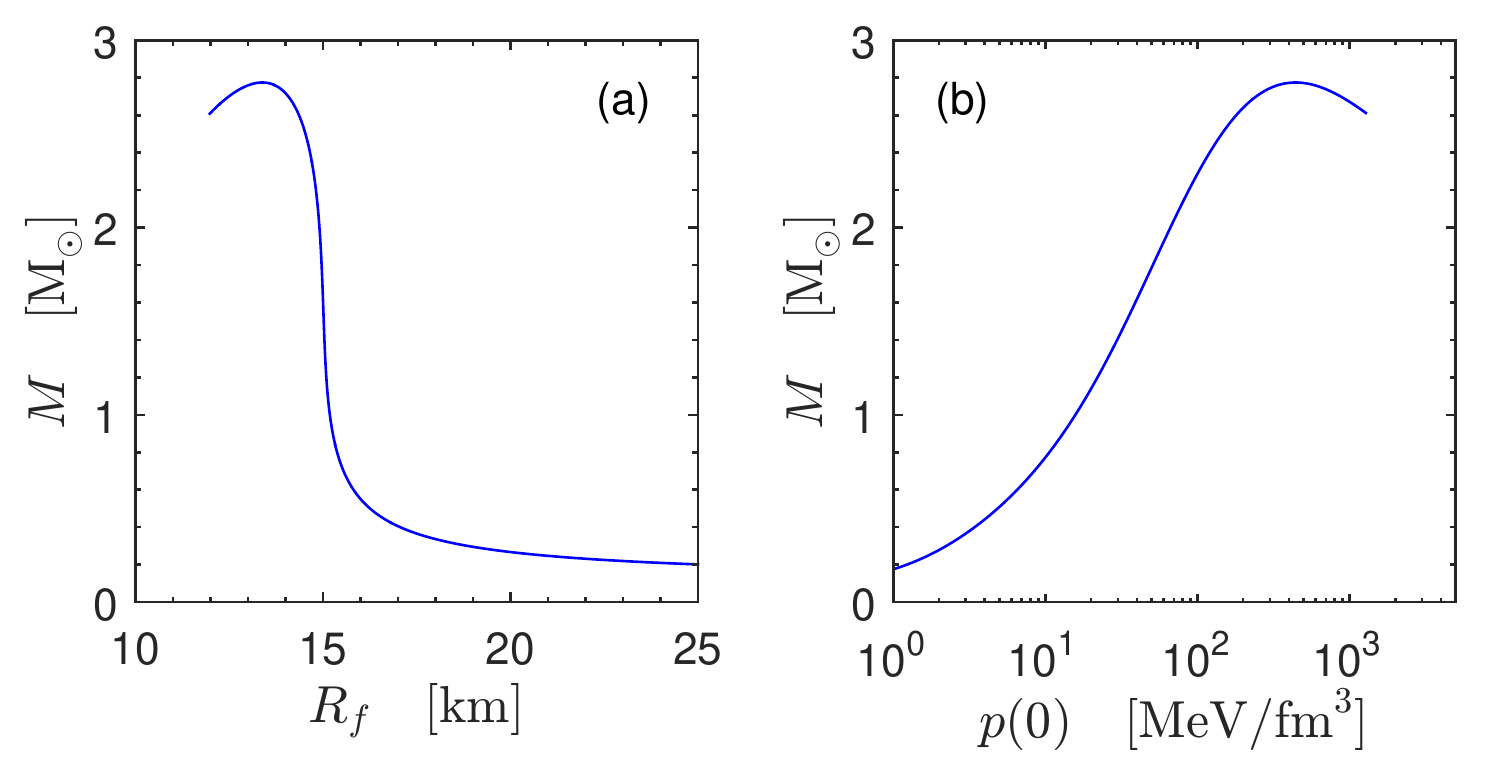}
\caption{(a) The mass\hyp radius curve and (b) the mass as a function of the central pressure for fermion stars with the NL3 equation of state.}
\label{fig:NL3}
\end{figure}

Some results are shown in Fig.\ \ref{fig:NL3}.  In Fig.\ \ref{fig:NL3}(a) we show the mass-radius curve and in Fig.\ \ref{fig:NL3}(b) we show the mass as a function of the central pressure.  It is well\hyp known that for a single\hyp fluid system, such as considered in this section, the transition from stable to unstable occurs at the solution with the largest mass \cite{GlendenningBook}, which is called the critical solution.  A solution with a central pressure smaller than the critical pressure is stable, otherwise it is unstable.  The critical solution (for the NL3 equation of state that we are using) has $p(0) = 441$ MeV/fm$^3$, $M = 2.78$ M$_\odot$, and $R_f = 13.4$ km.

\section{Charged boson stars}
\label{sec:CBS}

In this section, we review charged boson stars \cite{Jetzer:1989av}.  The equations derived here will be directly applicable to fermion--charged-boson stars.  We begin with the standard Lagrangian for a complex scalar field gauged under $U(1)$,
\begin{equation} \label{b Lagrangian}
\mathcal{L}_b
= -\left(D_\mu \phi \right) \left(D^\mu \phi \right)^* - V(|\phi|) - \frac{1}{4} F_{\mu\nu} F^{\mu\nu},
\end{equation}
where $\phi$ is the complex scalar field,
\begin{equation} \label{b Lag eqs}
\begin{split}
D_\mu \phi &= \partial_\mu \phi - ig A_\mu \phi \\
F_{\mu\nu} &= \partial_\mu A_\nu - \partial_\nu A_\mu \\
V &= \mu^2 |\phi|^2 + \lambda |\phi|^4,
\end{split}
\end{equation}
$A_\mu$ is the gauge field and is real, with $F_{\mu\nu}$ its field strength, $\mu$ is the scalar field particle mass, $\lambda$ is the self-coupling constant, and $g$ is the boson charge.  For completeness, we derive equations for arbitrary $\lambda$, but, for simplicity, present results only for $\lambda = 0$.  In spherically symmetric systems,
\begin{equation}
A_\mu = (A_t, A_r, 0 ,0),
\end{equation}
and the only nonzero components of the field strength are $F_{tr} = -F_{rt} = \partial_t A_r - \partial_r A_t$.  For convenience, we give the flat space equations of motion for the scalar and gauge fields,
\begin{equation} \label{flat eom}
\begin{split}
0 &=  D_\mu D^\mu \phi  - \frac{d V}{d|\phi|^2} \phi \\
0 &= \partial_\mu F^{\mu\nu} - J^{\nu},
\end{split}
\end{equation}
which are the (gauged) Klein-Gordon and Maxwell equations, and where 
\begin{equation} \label{flat J}
J^{\mu} =  ig \left[\phi^*D^\mu\phi - \phi \bigl(D^\mu \phi \bigr)^*\right]
\end{equation}
is the conserved matter current.  That it is conserved follows immediately from $F^{\mu\nu}$ being antisymmetric.  The Lagrangian in Eq.\ (\ref{b Lagrangian}) is invariant under a $U(1)$ transformation,
\begin{equation}
\phi \rightarrow \phi' = e^{-i\Lambda} \phi, \qquad
A_\mu \rightarrow A_\mu' = A_\mu - \frac{1}{g} \partial_\mu \Lambda,
\end{equation}
where $\Lambda$ is the gauge parameter.  This allows for some freedom in how we parametrize $\phi$ and $A_\mu$.  We shall make a particular gauge choice below that is convenient for finding static solutions.

We minimally couple the Lagrangian in Eq.\ (\ref{b Lagrangian}) to gravity via $\mathcal{L}_b \rightarrow \sqrt{-\det(g_{\mu\nu})} \mathcal{L}_b$, where $\det(g_{\mu\nu})$ is the determinant of the metric.  The energy\hyp momentum tensor is given by
\begin{align}
(T_f)\indices{^\mu_\nu} &= 
\left(D^\mu \phi \right) \left(D_\nu \phi \right)^* 
+ \left(D^\mu \phi \right)^* \left(D_\nu \phi \right)
\notag\\
&\qquad
-\delta^\mu_\nu
 \left(D_\sigma\phi\right)\left(D^\sigma\phi\right)^*
-\delta^\mu_\nu V
\notag\\
&\qquad
 + F^{\mu \sigma} F_{\nu \sigma}
- \frac{1}{4} \delta^\mu_\nu F_{\sigma\lambda}F^{\sigma\lambda}.
\end{align}
The equations of motion may be obtained directly from the Lagrangian $\sqrt{-\det(g_{\mu\nu})} \mathcal{L}_b$ or by elevating the derivatives in Eq.\ (\ref{flat eom}) to derivatives covariant with respect to gravity.  In writing the equations of motion and the components of the energy\hyp momentum tensor, it is convenient to define
\begin{equation} \label{q def}
\mathcal{Q} \equiv -\frac{4\pi r^2}{\alpha a} F_{tr} = -\frac{4\pi r^2}{\alpha a}
(\partial_t A_r - \partial_r A_t).
\end{equation}
We shall find below that $\mathcal{Q}$ gives the total charge inside a radius $r$.  We note that we have defined $F_{tr} \equiv \partial_t A_r - \partial_r A_t$, with lower indices, from which $F^{tr} = - F_{tr}/(\alpha^2 a^2)$.  The equations of motion are then
\begin{align} 
\partial_t \left( \frac{a}{\alpha} D_t \phi \right) &=
\frac{1}{r^2} \partial_r
\left(
 \frac{r^2 \alpha }{a} D_r \phi
\right)
+ig\left( \frac{a}{\alpha} A_t D_t \phi - \frac{\alpha }{a} A_r D_r \phi \right)
\notag \\
&\qquad
- \alpha a\frac{d V}{d|\phi|^2} \phi
\notag \\
\dot{\mathcal{Q}} &= +4\pi \alpha a  r^2 J^r
\notag \\
\mathcal{Q}' &= - 4\pi\alpha a r^2  J^t,
\label{b eom}
\end{align}
where $D_\mu \phi$ is as given in Eq.\ (\ref{b Lag eqs}) and
\begin{equation} \label{b current}
\begin{split}
J^t &=
-\frac{ig}{\alpha^2} \left[ \phi^* D_t\phi - \phi (D_t \phi)^* \right]
\\
J^r &= + \frac{ig}{a^2} \left[
\phi^* D_r\phi 
- \phi (D_r\phi)^* \right].
\end{split}
\end{equation}
The energy\hyp momentum tensor components that we will be using are 
\begin{equation} \label{T comps}
\begin{split}
(T_b)\indices{^t_t}
&= -\frac{1}{\alpha^2} |D_t\phi|^2 - \frac{1}{a^2}|D_r\phi|^2
- V
- \frac{\mathcal{Q}^2}{32\pi^2 r^4}
\\
(T_b)\indices{^r_r}
&=+\frac{1}{\alpha^2} |D_t\phi|^2 + \frac{1}{a^2}|D_r\phi|^2
 - V - \frac{ \mathcal{Q}^2}{32\pi^2 r^4}
\\
(T_b)\indices{^t_r}
&=
- \frac{1}{\alpha^2} \left[(D_t\phi) (D_r\phi)^* + (D_t\phi)^* D_r\phi \right].
\end{split}
\end{equation}

As mentioned, static solutions require the metric and energy-momentum tensor to be time independent.  A look at the $\dot{a}$ equation in (\ref{metric equations}) shows that we must also have $(T_b)\indices{^t_r} = 0$.  The best gauge to use when finding static solutions is radial gauge, for which
\begin{equation}
A_r = 0.
\end{equation}
For the energy\hyp momentum tensor to be time\hyp independent, it follows from Eqs.\ (\ref{T comps}) that $|D_t\phi|^2$ and $\mathcal{Q}$ are time\hyp independent, which leads to $A_t$ being time\hyp independent and $\phi = \phi(r) e^{-i \omega t}$ for complex $\phi(r)$ and real constant $\omega$.  A simple gauge transformation, with gauge parameter $\Lambda = \omega t$, removes the time\hyp dependence of $\phi$, while retaining $A_r = 0$ and shifting $A_t$ by a constant, which we absorb into $A_t$.  

A more complicated gauge transformation transforms $\phi$ to be purely real, as we now explain.  We have just established that $\phi = \phi(r)$.  This can be written in terms of real and imaginary parts or in terms of a modulus and phase,
\begin{equation}
\phi(r) = \phi_r(r) + i\phi_i(r) = |\phi(r)| e^{i\tan^{-1}[\phi_i(r)/\phi_r(r)]}.
\end{equation}
Making a gauge transformation with gauge parameter $\Lambda = -\tan^{-1}(\phi_i/\phi_r)$ transforms $\phi$ to be purely real.  Since $\phi_r,\phi_i$ are time\hyp independent, $\partial_t \Lambda = 0$ and $A_t$ is unchanged.  It is also the case that $\partial_r \Lambda = 0$ and we stay in radial gauge with $A_r = 0$.  To show this, we have from $\dot{\mathcal{Q}} = 0$ in Eq.\ (\ref{b eom}) that $J^r = 0$ and thus that $\phi^*D_r\phi = \phi (D_r\phi)^*$ which, with $A_r = 0$, becomes $\phi_r \phi_i' = \phi_r' \phi_i$.  This is the necessary result to show that both $\partial_r \Lambda = 0$ and $(T_b)\indices{^t_r} = 0$.

By making convenient gauge choices, we have that all fields (both metric and matter fields) are time\hyp independent, that $\phi$ is real, and that $A_r = 0$.  When presenting results, it is best to present gauge\hyp independent quantities, which are $|\phi|$ and $F_{tr}$.  From (\ref{q def}), $\mathcal{Q}$ is also gauge\hyp independent.  The equations for the neutral boson star are obtained in the limit $g\rightarrow 0$.  Since our gauge choice has absorbed a constant into $A_t$ that is proportional to $1/g$, this limit is more easily taken after separating out the constant.  Numerically, we can simply set $g$ to a small number, which we have confirmed can accurately reproduce results for a neutral boson star.

In writing the final form of the static equations, we move to the metric variables $\sigma$ and $m$ defined in Eq.\ (\ref{sigma m def}).  The equations of motion are then
\begin{align} \label{boson star eom}
\phi'' &=
- \left\{\frac{2}{r}
+ \frac{4\pi G r}{N} \left[ (T_\text{tot})\indices{^r_r} + (T_\text{tot})\indices{^t_t} \right] + \frac{2 G m}{Nr^2}
 \right\} \phi'
\notag\\
&\qquad
- \frac{1}{N} \left( \frac{ g^2 A_t^2}{N\sigma^2} - \mu^2 - 2\lambda \phi^2 \right) \phi
\notag\\
A_t''
&= 
\left\{ \frac{4\pi G r}{N} \left[ (T_\text{tot})\indices{^r_r} - (T_\text{tot})\indices{^t_t} \right] - \frac{2}{r} \right\} A_t'
+ \frac{2g^2}{N} A_t \phi^2,
\end{align}
where $N = 1 - 2G m/r$, and the energy\hyp tensor components are
\begin{equation} \label{Tb}
\begin{split}
(T_b)\indices{^t_t}  &= -N \phi^{\prime\, 2} - \frac{g^2 A_t^2 \phi^2}{N \sigma^2}
- \mu^2 \phi^2 - \lambda \phi^4
- \frac{A_t^{\prime \, 2}}{2 \sigma^2}
\\
(T_b)\indices{^r_r}  &= +N \phi^{\prime\, 2} + \frac{g^2 A_t^2 \phi^2}{N \sigma^2} 
- \mu^2 \phi^2 - \lambda \phi^4
- \frac{A_t^{\prime \, 2}}{2 \sigma^2}.
\end{split}
\end{equation}
Charged boson stars are given by the solutions to Eqs.\ (\ref{boson star eom}) and (\ref{sigma m eqs}) with $(T_\text{tot})\indices{^\mu_\nu} = (T_b)\indices{^\mu_\nu}$.  To solve these equations, we need inner and outer boundary conditions.  For inner boundary conditions we require that the energy\hyp momentum tensor be finite at the origin and thus $\mathcal{Q}(r) = O(r^2)$.  Additional inner boundary conditions are found by plugging Taylor expansions of the fields into the equations of motion.  We find $\phi(r) = \phi(0) + O(r^2)$, $A_t(r) = A_t(0) + O(r^2)$, $\sigma(r) = \sigma(0) + O(r^2)$ and $m = O(r^3)$.  It helps with finding solutions to define
\begin{equation}
\hat{A}_t(r) \equiv \frac{A_t(r)}{\sigma (0)}, \qquad
\hat{\sigma}(r) \equiv \frac{\sigma(r)}{\sigma (0)}.
\end{equation}
After replacing $A_t$ and $\sigma$ with $\hat{A}_t$ and $\hat{\sigma}$, $\sigma(0)$ cancels out.  Further, the inner boundary conditions are $\hat{A}_t(r) = \hat{A}_t(0) + O(r^2)$ and $\hat{\sigma}(r) = 1 + O(r^2)$ and $\sigma(0)$ no longer has to be known.  How we determine $\hat{A}_t(0)$ will be explained in the next paragraph.  To obtain outer boundary conditions, we require the energy\hyp momentum tensor to go to zero at large $r$.  We thus have $\phi,\phi',A_t' \rightarrow 0$ as $r\rightarrow \infty$.

Static solutions are uniquely identified by the central value $\phi(0)$.  We thus specify a value for $\phi(0)$ and a trial value for $\hat{A}_t(0)$.  We write the equations of motion in (\ref{boson star eom}) in first order form and then integrate Eqs.\ (\ref{boson star eom}) and (\ref{sigma m eqs}) outward from $r = 0$ using the previously listed inner boundary conditions.  We then vary $\hat{A}_t(0)$ using the shooting method until the outer boundary conditions are satisfied, at which point we have found a static solution.

Before presenting results, there are three additional quantities that are needed.  These are equations for the total charge and mass of the star and our definition for the star's radius.  As mentioned, the current in Eq.\ (\ref{b current}) is conserved, $\nabla_\mu J^\mu = 0$, which describes charge conservation.  The total charge inside a radius $r$ is given by
\begin{equation}
-\int_0^r d^3r \sqrt{-g} J^t.
\end{equation} 
Using the bottom equation in (\ref{b eom}), this can be integrated exactly and equals $\mathcal{Q}(r)$.  Thus, as promised, $\mathcal{Q}(r)$ gives the total charge inside a radius $r$.  The total charge in the system is given by $Q = \mathcal{Q}(\infty)$.  Note that once a static solution is found, $\mathcal{Q}(r)$ is determined algebraically from Eq.\ (\ref{q def}), 
\begin{equation}
\mathcal{Q}(r) = 4\pi r^2 \frac{\hat{A}_t'(r)}{\hat{\sigma}(r)},
\end{equation}
where both $\hat{A}_t'$ and $\hat{\sigma}$ follow from the static solution.

To find an equation for the total mass of the star, we require that the spacetime be asymptotically Reissner-Nordstr\"om, so that in the large $r$ limit
\begin{equation}
\frac{1}{a^2} \rightarrow 1 - \frac{2 G M}{r} + \frac{ G (Q^2/4\pi)}{r^2},
\end{equation}
which defines $M$, which we take to be the mass of the star.  Comparing this to the definition of $m(r)$ in Eq.\ (\ref{sigma m def}), we define
\begin{equation} \label{script M}
\mathcal{M}(r) \equiv m(r) + \frac{\mathcal{Q}^2(r)}{8\pi r},
\end{equation}
so that $M = \mathcal{M}(\infty)$.

In the study of neutral boson stars, it is customary to define the radius of the star as the radius that contains some large percentage of the mass of the star.  The same can be done with charged boson stars.  We take the radius of the star to be $R_{95}$, which is the radius that contains 95\% of the mass and is defined by
\begin{equation}
\mathcal{M}(R_{95}) = (0.95) M.
\end{equation}

In presenting results, we use the quantities
\begin{equation}
\begin{split}
\overline{M} \equiv \left( \frac{\mu}{10^{-10} \text{ eV}}  \right) M,
\qquad
\overline{R}_{95} \equiv \left( \frac{\mu}{10^{-10} \text{ eV}}  \right) R_{95},
\end{split}
\end{equation}
which will be given in solar masses and kilometers, and 
\begin{equation}
|\bar{\phi}(0)| \equiv \frac{1}{m_P}|\phi(0)|,
\qquad
\bar{g} \equiv \frac{m_P}{\sqrt{8\pi}  \mu} g,
\end{equation}
which are dimensionless, where $m_P = 1/\sqrt{G}$ is the Planck mass.  The factor of $\sqrt{8\pi}$ is included in the definition of $\bar{g}$ so that it agrees with the scaled charge used in the literature \cite{Jetzer:1989av, Pugliese:2013gsa}.  A nonrelativistic analysis suggests solutions only exist for $\bar{g}^2 < 1/2$, since otherwise the Coulomb repulsion is greater than the gravitational attraction \cite{Jetzer:1989av}.  This is not necessarily true in the relativistic case \cite{Pugliese:2013gsa}.  Nevertheless, we consider only $\bar{g} < 1/\sqrt{2} = 0.7071$.

\begin{figure}
\centering
\includegraphics[width=2.85in]{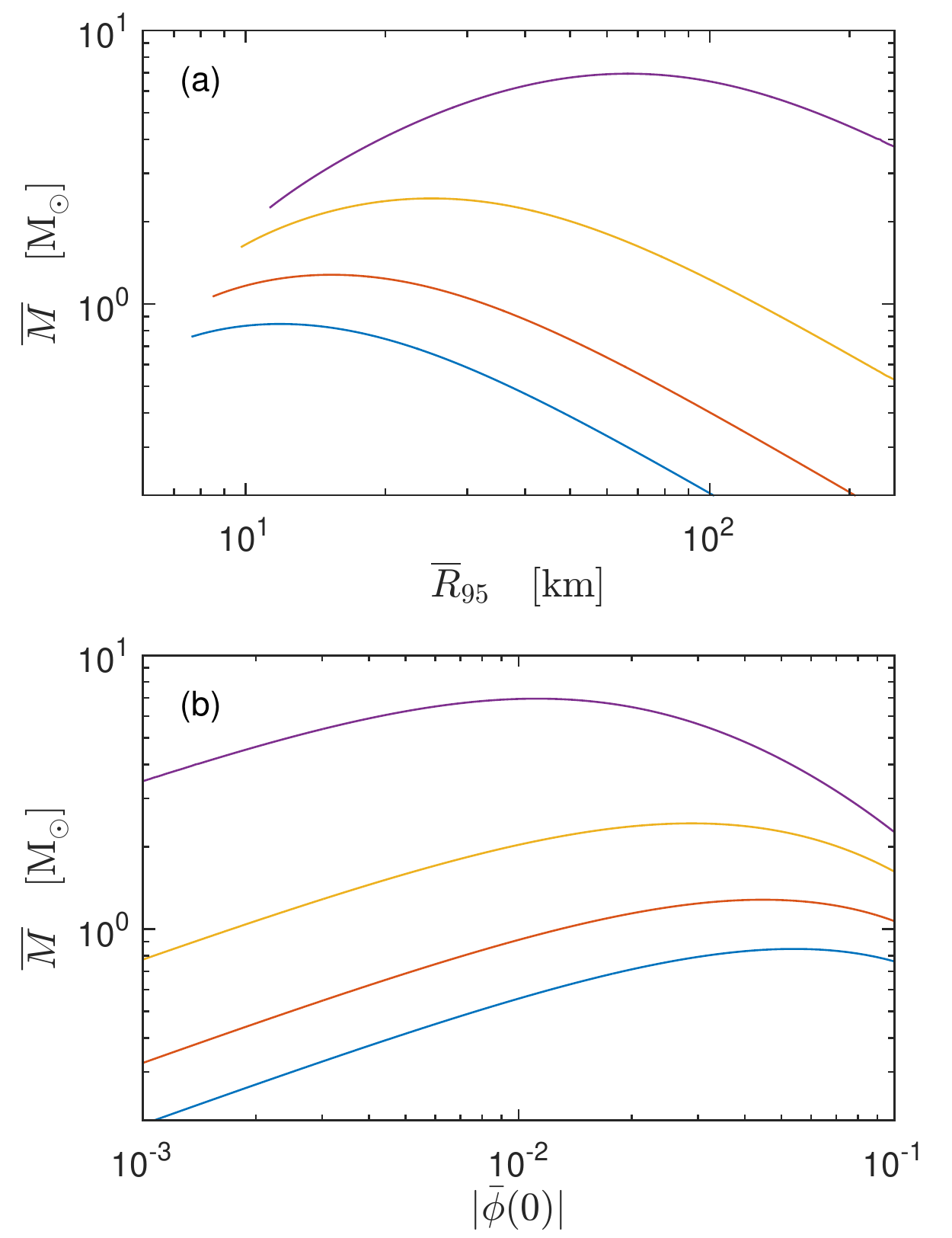}
\caption{(a) Mass\hyp radius curves and (b) the mass as a function of the central value of the scalar field for charged boson stars.  In both plots, the curves, from bottom to top, are for $\bar{g} = 0$ (blue), 0.5 (red), 0.65 (orange), and 0.7 (purple).}
\label{fig:CBS}
\end{figure}

Some results are shown in Fig.\ \ref{fig:CBS}.  In Fig.\ \ref{fig:CBS}(a) we show mass-radius curves for various values of $\bar{g}$.  In Fig.\ \ref{fig:CBS}(b) we show the mass as a function of the central (gauge\hyp independent) value $|\bar{\phi}(0)|$.  Many additional plots can be found in \cite{Pugliese:2013gsa}.  The transition from stable to unstable occurs at the critical solution.  Since charged boson stars are single\hyp fluid systems, the critical solution is the solution with the largest mass, just as with fermion stars.  Solutions with central values of $|\phi(0)|$ smaller than the critical value are stable, otherwise they are unstable.  The critical values of $(|\bar{\phi}(0)|, M, R_{95})$ are (0.0541, 0.846 M$_\odot$, 11.9 km) for $\bar{g} = 0$, (0.0447, 1.28 M$_\odot$, 15.3 km) for $\bar{g} = 0.5$, (0.0288, 2.43 M$_\odot$, 25.1 km) for $\bar{g} = 0.65$, and (0.0112, 6.95 M$_\odot$, 66.7 km) for $\bar{g} = 0.7$.  We can see that the critical solution can have its mass and radius increased significantly by increasing $\bar{g}$ \cite{Jetzer:1989av}.  

\section{Fermion--charged-boson stars}
\label{sec:FCBS}

In this section, we combine the equations from the previous sections and study fermion--charged-boson stars.  As mentioned in the Introduction, we assume that nongravitational interactions between the fermion and boson sectors are negligible.  Consequently, the energy\hyp momentum tensor separates \cite{Henriques:1989ar},
\begin{equation}
(T_\text{tot})\indices{^\mu_\nu} = (T_f)\indices{^\mu_\nu} + (T_b)\indices{^\mu_\nu},
\end{equation}
where $(T_f)\indices{^\mu_\nu}$ and $(T_b)\indices{^\mu_\nu}$ are as given in Eqs.\ (\ref{Tf}) and (\ref{Tb}) and are independently conserved,
\begin{equation}
\nabla_\mu (T_f)\indices{^\mu_\nu} = 0, \qquad \nabla_\mu (T_b)\indices{^\mu_\nu} = 0,
\end{equation}
which leads to the equations of motion given in Eqs.\ (\ref{f eom}) and (\ref{boson star eom}).  We have also the metric equations in (\ref{sigma m eqs}) and (\ref{alpha prime eq}).  The inner and outer boundary conditions are unchanged from those given in Secs.\ \ref{sec:FS} and \ref{sec:CBS}.

Static solutions are uniquely specified by the central pressure $p(0)$ for the fermion sector and the central value $\phi(0)$ for the boson sector.  To solve the full system of equations, we specify values for $p(0)$ and $\phi(0)$ and a trial value for $\hat{A}_t(0)$.  We then integrate outward from $r=0$ and vary $\hat{A}_t(0)$ using the shooting method until the outer boundary conditions for the boson sector are satisfied, at which point we have found a static solution.  During the integration, either $p$ hits zero or $\phi$ drops to such a low value that it is effectively zero.  When this happens, we break the integration and then restart it using only the single\hyp fluid equations from either Sec.\ \ref{sec:FS} or Sec.\ \ref{sec:CBS}.  The radius of the fermion part of the star, $R_f$, is defined as the radial position where the pressure hits zero, $p(R_f) = 0$.  We consider this the visible radius, since it is the ordinary fermionic matter, and not dark matter, that is visible.  The mass of the total system, $M$, is found using Eq.\ (\ref{script M}), but now $m(r)$ has contributions from both the fermion and boson sectors.


\subsection{Stability}

The first property we study is the stability of the static solutions with respect to small perturbations.  In single\hyp fluid systems, this is straightforward, since the transition from stable to unstable occurs at the static solution with the largest mass.  As explained in Secs.\ \ref{sec:FS} and \ref{sec:CBS}, this solution is called the critical solution and it is identified by the critical central pressure, $p(0)$, for fermion stars and the critical central value for the scalar field, $|\phi(0)|$, for charged boson stars.  The situation is not as straightforward for fermion--charged-boson stars, since static solutions are identified by both $p(0)$ and $|\phi(0)|$.  As such, stability is no longer marked by a critical point, but instead a critical curve.  One possibility for computing the critical curve is to write each field as a perturbation about its static value and then to solve the linearized system of equations.  This has been done for fermion stars \cite{Chandrasekhar:1964zz}, boson stars \cite{Gleiser:1988rq, Jetzer:1988vr, Gleiser:1988ih, Lee:1988av, Hawley:2000dt, kain}, and charged boson stars \cite{Jetzer:1989us}, but not, as far as we are aware, for fermion-boson stars.

A simpler method for computing critical curves was presented in \cite{Henriques:1990xg}.  Critical curves satisfy
\begin{equation} \label{critical curve}
\frac{d M}{d\mathbf{p}} = \frac{d N_f}{d\mathbf{p}} = \frac{d N_b}{d\mathbf{p}} = 0,
\end{equation}
where $M$ is the mass of the system, $N_f$ and $N_b$ are the conserved fermionic and bosonic particle numbers, and $\mathbf{p}$ is a vector in parameter space.  For details on the derivation of this result, we refer the reader to \cite{Henriques:1990xg, ValdezAlvarado:2012xc, Kain:2021hpk}.  It can be shown that if two of the quantities in Eq.\ (\ref{critical curve}) are zero, then the third is also \cite{Henriques:1990xg, Jetzer:1990xa}.  One thing we note is that if we were to compute the critical curve by perturbing the system, we would only consider perturbations that conserve $M$, $N_f$, and $N_b$ \cite{Chandrasekhar:1964zz, Gleiser:1988rq, Jetzer:1988vr, Gleiser:1988ih}.  This is the reason why these quantities show up in Eq.\ (\ref{critical curve}).

Before explaining how we solve Eq.\ (\ref{critical curve}), we review how $N_f$ and $N_b$ are computed.  It is worth stressing that $N_f$ is the number of fluid elements and not, say, the number of baryons.  The number density of fluid elements for the fermion sector, $n_f$, can be computed directly from the equation of state using \cite{Kain:2021hpk}
\begin{equation}
n_f \propto \exp\left(\int \frac{d\epsilon}{\epsilon + p} \right), 
\end{equation}
where the proportionality constant does not affect the determination of stability and therefore does not have to be known.  The total number of fermionic fluid elements inside a radius $r$, $\mathcal{N}_f(r)$, obeys
\begin{equation}
\mathcal{N}_f' = \frac{4\pi r^2 n_f}{\sqrt{N}},
\end{equation}
where $N = 1 - 2Gm/r$, and $N_f = \mathcal{N}_f(\infty)$.  The equation above can be solved simultaneously with the equations of motion and metric equations.  For the boson sector, $N_b$ is the conserved particle number and is proportional to the conserved charge.  Defining $\mathcal{N}_b(r) \equiv \mathcal{Q}(r)/g$, the large $r$ limit of $\mathcal{N}_b$ gives $N_b$, or
\begin{equation}
N_b = \frac{Q}{g}.
\end{equation}

To solve Eq.\ (\ref{critical curve}) we follow the methods presented in \cite{ValdezAlvarado:2012xc, Kain:2021hpk}.  We compute contour lines of either $N_f$ or $N_b$ in the full system.  Moving along a single contour line, we determine the point where $M$ is an extremum (in practice, we find that it is a maximum).  These points give the critical curve.

\begin{figure*}
\centering
\includegraphics[width=6.5in]{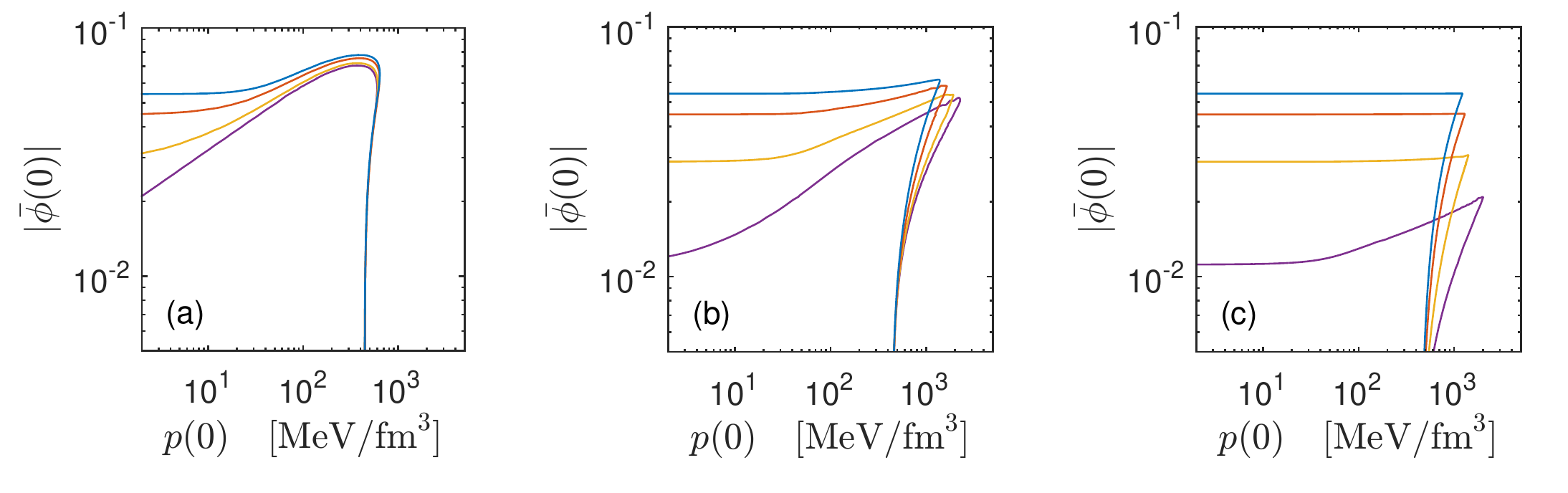}
\caption{Critical curves for (a) $\mu = 10^{-10}$ eV, (b) $10^{-9}$ eV, and (c) $10^{-8}$ eV.  A point in each plot represents a static solution describing a fermion--charged-boson star.  Those points inside the critical curve (i.e.\ the lower left region enclosed by the curve) represent stable solutions.  In each plot, the curves, from top to bottom, are for $\bar{g} = 0$ (blue), 0.5 (red), 0.65 (orange), and 0.7 (purple).}
\label{fig:crit_cruves}
\end{figure*}

Critical curves for $\mu = 10^{-10}$ eV and various values of $\bar{g}$ are shown in Fig.\ \ref{fig:crit_cruves}(a).  Each point in Fig.\ \ref{fig:crit_cruves}(a) represents a static solution.  Those points that are inside the curve (i.e.\ the lower left region enclosed by the curve) represent stable solutions.  For sufficiently small $|\phi(0)|$, we expect the star to be dominated by the fermion sector.  Similarly, for sufficiently small $p(0)$, we expect the star to be dominated by the boson sector.  In Fig.\ \ref{fig:crit_cruves}(a), we can see that for sufficiently small $|\phi(0)|$ or $p(0)$, the critical curves are trending towards, if not reproducing, the critical values reviewed in Secs.\ \ref{sec:FS} and \ref{sec:CBS}, as expected.  The most interesting region of parameter space is when both $p(0)$ and  $|\phi(0)|$ are relativity large, which occurs at the upper\hyp right corner of the curves.  Here we find that stability can be increased beyond the single\hyp fluid critical values from Secs.\ \ref{sec:FS} and \ref{sec:CBS}.  Indeed, we find that for larger $\bar{g}$, stability can be increased well beyond the single\hyp fluid critical values of $|\bar{\phi}(0)|$.  The critical curves shown in Fig.\ \ref{fig:crit_cruves}(a) are similar in shape to the critical curves for dark matter admixed neutron stars with fermionic dark matter \cite{Kain:2021hpk}.

Figures \ref{fig:crit_cruves}(b) and \ref{fig:crit_cruves}(c) show, respectively, critical curves for $\mu = 10^{-9}$ eV and $\mu = 10^{-8}$ eV.  As the scalar field particle mass is increased we can see the extent to which the critical curve extends beyond the single\hyp fluid critical value for $|\bar{\phi}(0)|$ is decreased.  This can only occur if the fermion sector has a smaller influence over stability.  We see also that the critical curve extends further beyond the single\hyp fluid critical value for $p(0)$, indicating that the boson sector is having a larger influence over stability.  We thus find that an increase in the scalar field mass leads to an increase in the role played by the boson sector for stability.


\subsection{Fermion and boson contributions}

It is useful to determine the relative contributions from the fermion and boson sectors to the star.  In the original work on fermion-boson stars \cite{Henriques:1989ar}, this was studied in terms of the relative particle numbers.  We choose instead to look at the relative contributions to the total mass.  Since the mass of the star is measurable, it is interesting to determine how much the fermion and boson sectors contribute to such a measurement.

The solution to the bottom equation in (\ref{sigma m eqs}) is $m$, which is used to find the mass of the star.  This equation conveniently splits into $m = m_f + m_b$, where
\begin{equation}
m_f' \equiv -4\pi r^2 (T_f)\indices{^t_t}, \qquad m_b' \equiv -4\pi r^2 (T_b)\indices{^t_t}.
\end{equation}
From Eq.\ (\ref{script M}), we then have $\mathcal{M} = \mathcal{M}_f + \mathcal{M}_b$, where
\begin{equation}
\mathcal{M}_f \equiv m_f, \qquad
\mathcal{M}_b \equiv m_b + \frac{\mathcal{Q}^2}{8\pi r}.
\end{equation}
The total mass of the star can then be written as  $M = M_f + M_b$, with the fermion and boson sector contributions given by
\begin{equation}
M_f \equiv \mathcal{M}_f(\infty), \qquad
M_b \equiv \mathcal{M}_b(\infty).
\end{equation}

\begin{figure*}
\centering
\includegraphics[width=7in]{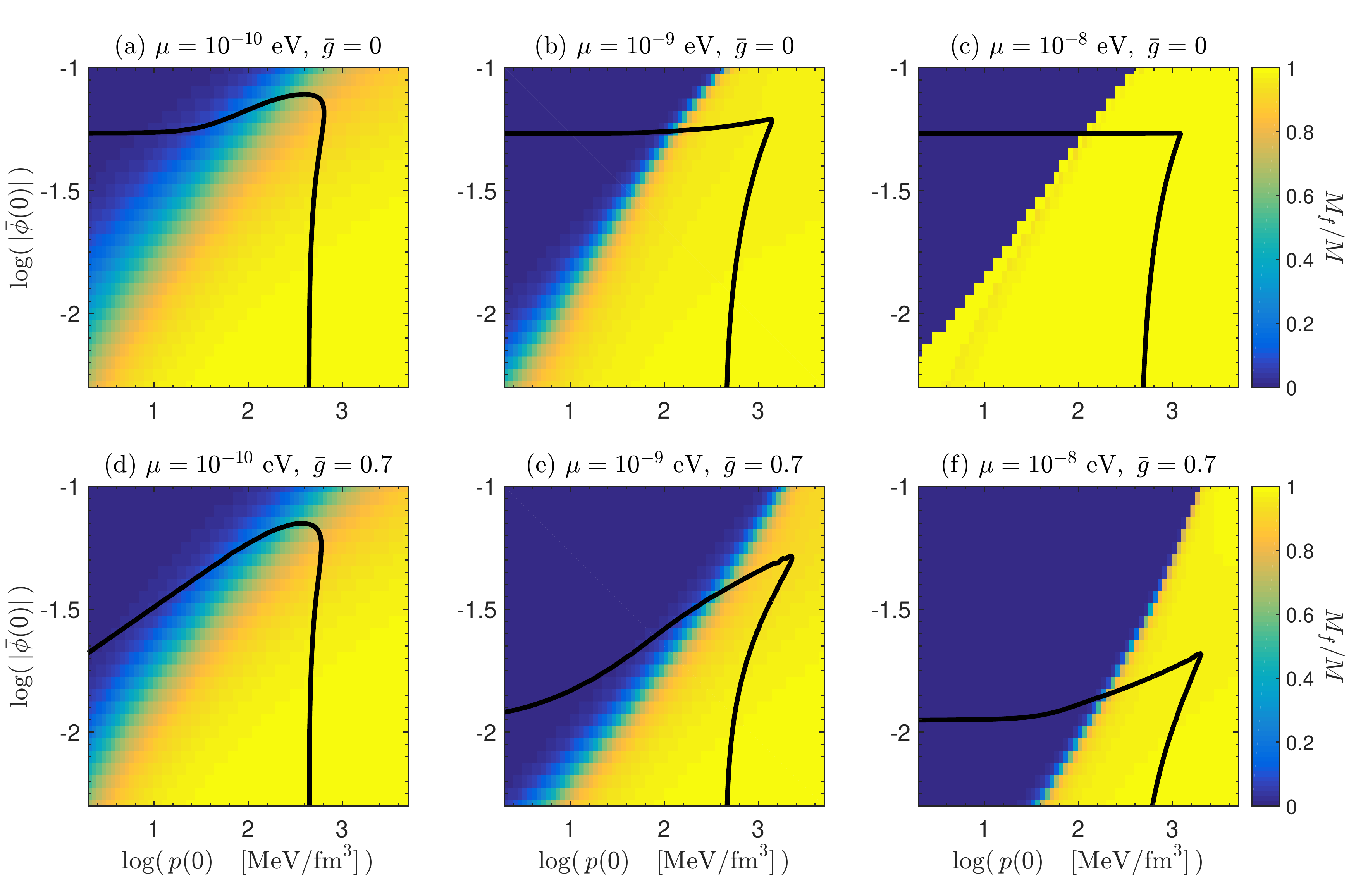}
\caption{The contribution from the fermion sector to the total mass of the star, $M_f/M$, is shown for various values of the scalar field mass, $\mu$, and the scalar field charge, $\bar{g}$.  These plots show the relative contribution to the total mass of the star from the boson and fermion sectors.  Bright yellow indicates that the mass of the star is fermion dominated and dark blue indicates that it is boson dominated.  The thick black curves are the critical curves from Fig.\ \ref{fig:crit_cruves}.}
\label{fig:crit_contributions}
\end{figure*}

In Fig.\ \ref{fig:crit_contributions}, we show density plots for $M_f/M$, which is the contribution from the fermion sector to the total mass of the star.  The upper row is for $\bar{g} = 0$ and for (a) $\mu = 10^{-10}$ eV, (b) $10^{-9}$ eV, and (c) $10^{-8}$ eV.  The bottom row plots the same, but for $\bar{g} = 0.7$.  Consider first the left column, where Figs.\ \ref{fig:crit_contributions}(a) and \ref{fig:crit_contributions}(d) have $\mu = 10^{-10}$ eV.  We see a relativity wide transition from fermion domination (yellow) to boson domination (blue).  In other words, the star is well mixed with fermions and bosons.  This is exactly what we expected for $\mu = 10^{-10}$ eV from the simple analysis done in the Introduction.  On the other hand, we see in the right column, where Figs.\ \ref{fig:crit_contributions}(c) and \ref{fig:crit_contributions}(f) have $\mu = 10^{-8}$ eV, that the transition is quite sharp.  That the transition between fermion and boson domination sharpens as the scalar field mass is increased was first found in \cite{Henriques:1989ar} for fermion--neutral-boson stars.  Here we show that it continues to hold with charged scalar fields.

Superimposed on the plots in Fig.\ \ref{fig:crit_contributions} are the critical curves from Fig.\ \ref{fig:crit_cruves}.  Looking again at the right column, the yellow region is solidly dominated by the fermion sector, at least from the perspective of the total mass of the star.  However, if the star were truly a single\hyp fluid fermion star, then the vertical portion on the right side of critical curve would be straight.  That the vertical portion on the right side bends outward indicates that the boson sector still has influence over the star, even though it has a negligible contribution to the total mass of the star.  


\subsection{Mass-radius diagrams}

The last property we study are mass-radius diagrams.  Once a static solution is found, the mass of the total system, $M$, and the radius of the fermion sector, $R_f$, are found as described at the beginning of this section.    We present results in terms of $R_f$, because $R_f$ is the visible radius of the star since it is the ordinary fermionic matter that is the visible matter.  Further, we present results only for $\mu = 10^{-10}$ eV.  

\begin{figure*}
\centering
\includegraphics[width=7in]{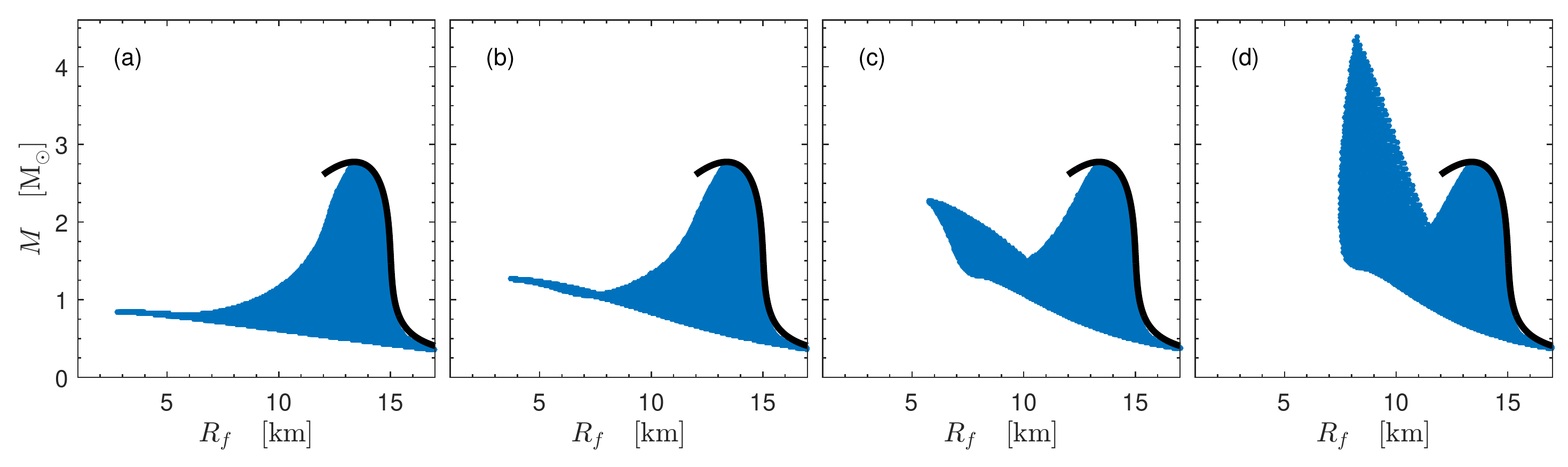}
\caption{Mass\hyp radius diagrams for fermion--charged-boson stars with $\mu = 10^{-10}$ eV and (a) $\bar{g} = 0$, (b) 0.5, (c) 0.65, and (d) 0.7.  $M$ is the total mass of the star and $R_f$ is the radius of the fermion sector, which is the visible radius of the star.  The blue region gives the mass\hyp radius relationships for the entirety of the stable parameter space shown in Fig.\ \ref{fig:crit_cruves}(a).  For comparison, the thick black curve is included, which is the mass-radius curve in Fig.\ \ref{fig:CBS}(a), i.e.\ the mass\hyp radius curve for a neutron star in the absence of dark matter.}
\label{fig:MRf}
\end{figure*}

In single\hyp fluid systems, one has mass\hyp radius curves, as shown in Figs.\ \ref{fig:NL3} and \ref{fig:CBS}.  With two\hyp fluid fermion--charged-boson stars, the mass\hyp radius relationship can no longer be represented by a single curve.  In Fig.\ \ref{fig:MRf}, we show mass\hyp radius diagrams for the entirety of the stable parameter space shown in Fig.\ \ref{fig:crit_cruves}(a).  In other words, for every stable point in Fig.\ \ref{fig:crit_cruves}(a), we compute the total mass and the radius of the fermion sector and plot this in Fig.\ \ref{fig:MRf}.  Figure \ref{fig:MRf} is analogous to the single\hyp fluid plots in Figs.\ \ref{fig:NL3}(a) and \ref{fig:CBS}(a).  For comparison, we have included the thick black curve, which is the mass-radius curve in Fig.\ \ref{fig:CBS}(a), i.e.\ the mass\hyp radius curve for a neutron star in the absence of dark matter.

We can see from Fig.\ \ref{fig:MRf} that the presence of the charged scalar field tends to decrease the visible radius of the star.  As the charge of the scalar field is increased, this leads to larger possible masses.  Thus, we find that fermion--charged-boson stars can be more compact and more massive than ordinary neutron stars.

\begin{figure*}
\centering
\includegraphics[width=7in]{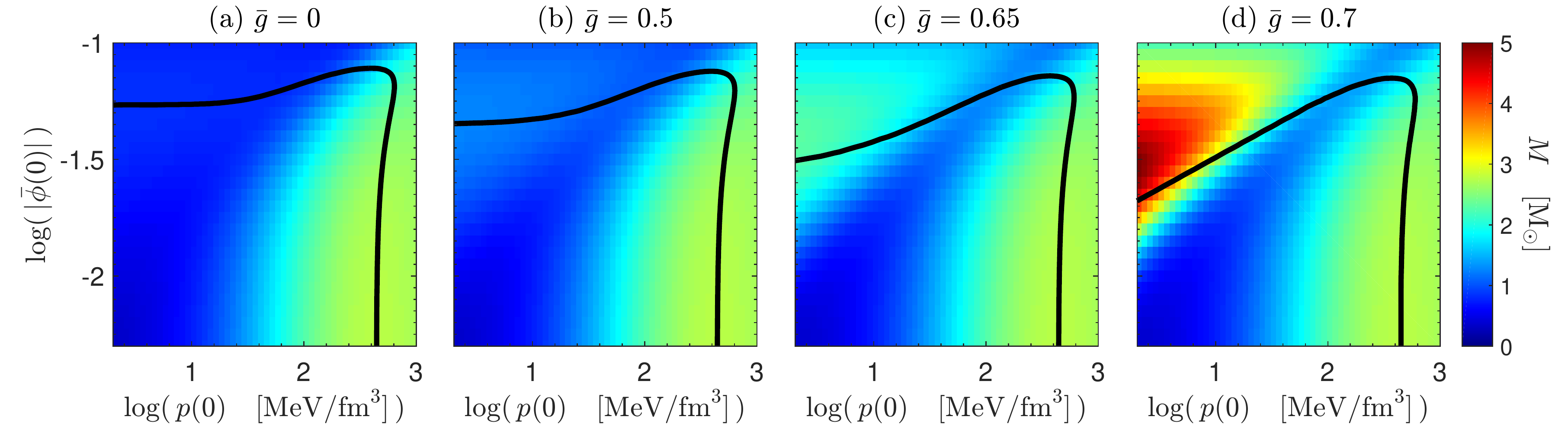}
\caption{Mass diagrams for fermion--charged-boson stars with $\mu = 10^{-10}$ eV and (a) $\bar{g} = 0$, (b) 0.5, (c) 0.65, and (d) 0.7.  The thick black curves are the critical curves from Fig.\ \ref{fig:crit_cruves}(a)}
\label{fig:mass_density}
\end{figure*}

In Fig.\ \ref{fig:mass_density}, we show mass diagrams, which are analogous to the single\hyp fluid diagrams in Figs.\ \ref{fig:NL3}(b) and \ref{fig:CBS}(b).  The thick black lines are the critical curves from Fig.\ \ref{fig:crit_cruves}(a).  We can see that for sufficiently small $|\phi(0)|$ or $p(0)$, the total mass of the system is given by the single\hyp fluid values, since in those regions the system is dominated by the fermion or boson sectors.  This is consistent with what we have seen in the previous two subsections.

\section{Conclusion}
\label{sec:conclusion}

We studied fermion--charged-boson stars, which are starlike systems with a mixture of fermions, as described by a perfect\hyp fluid energy\hyp momentum tensor and an equation of state, and bosons, as described by a charged complex scalar field.  Fermion--charged-boson stars are examples of dark matter admixed neutron stars with bosonic dark matter.

In the fermion sector, we used the NL3 equation of state, which is a realistic equation of state for the ordinary nuclear matter in a neutron star, and in the boson sector, we focused on a scalar field mass of $\mu = 10^{-10}$ eV.  We computed the critical curves, which show the stable region of parameter space.  We then studied how the fermion and boson sectors contribute to the total mass of the star, finding that the transition sharpens as the scalar field mass is increased.  Finally, we presented mass\hyp radius and mass diagrams, finding that charged bosonic dark matter can lead to more massive and more compact stars as the charge is increased.


\appendix

\section{NL3 equation of state}
\label{sec:NL3}

In this appendix, we review the construction of the NL3 equation of state \cite{Lalazissis:1996rd, ToddRutel:2005zz}.  The NL3 equation of state was used in the fermion sector throughout this paper and is constructed using relativistic mean field (RMF) theory.  For matter content, we include protons, neutrons, electrons, and muons, as well as the scalar, vector, and vector-isovector mesons.  We begin with a standard RMF theory Lagrangian \cite{GlendenningBook},
\begin{equation} \label{L ns}
\mathcal{L}_\text{ns} = \mathcal{L}_B + \mathcal{L}_\sigma + \mathcal{L}_\omega + \mathcal{L}_\rho + \mathcal{L}_L,
\end{equation}
where
\begin{equation}
\begin{split}
\mathcal{L}_B &= \sum_{z = p,n}
\bar{\psi}_z \biggl(i \gamma^\alpha \partial_\alpha - m_B + g_{B \sigma} \sigma 
\\
&\qquad\qquad - g_{B \omega} \gamma^\alpha \omega_\alpha - \frac{1}{2} g_{B \rho} \gamma^\alpha \vec{\rho}_\alpha \cdot \vec{\tau} \biggr) \psi_z
\\
\mathcal{L}_\sigma &= \frac{1}{2} \partial^\alpha \sigma \partial_\alpha \sigma - \frac{1}{2} m_\sigma^2 \sigma^2 
 - \frac{\kappa}{3!}  (g_{B \sigma} \sigma)^3 - \frac{\lambda}{4!} (g_{B \sigma } \sigma)^4
\\
\mathcal{L}_\omega &= -\frac{1}{4} \omega^{\alpha\beta} \omega_{\alpha\beta} + \frac{1}{2} m_\omega^2 \omega^\alpha \omega_\alpha
\\
\mathcal{L}_\rho &= -\frac{1}{4} \vec{\rho}^{\, \alpha\beta} \cdot \vec{\rho}_{ \alpha\beta} + \frac{1}{2} m_\rho \vec{\rho}^{\, \alpha} \cdot \vec{\rho}_\alpha
\\
\mathcal{L}_L &= \sum_{z=e,\mu} \bar{\psi}_z \left( i\gamma^\alpha \partial_\alpha - m_z \right) \psi_z,
\end{split}
\end{equation}
and 
\begin{equation}
\begin{split}
\omega_{\alpha\beta} &= \partial_\alpha \omega_\beta - \partial_\beta \omega_\alpha
\\
\vec{\rho}_{\alpha\beta} & = \partial_\alpha \vec{\rho}_\beta - \partial_\beta \vec{\rho}_\alpha - g_{\rho} (\vec{\rho}_\alpha \times \vec{\rho}_\beta).
\end{split}
\end{equation}
$\psi_z$, with $z= p,n,e,\mu$, are the wave functions for the proton, neutron, electron, and muon, $B$ labels the common baryon mass and couplings, $\sigma$ is the scalar meson, $v_\alpha$ is the vector meson, and $\vec{\rho}_\alpha$ is the vector-isovector meson.  We use the NL3 parameter set to fix the constants \cite{Lalazissis:1996rd, ToddRutel:2005zz}:
\begin{equation}
\begin{split}
m_B &= 938 \text{ MeV}
\\
m_\sigma &= 508.194 \text{ MeV}
\\
m_\omega &=782.5  \text{ MeV}
\\
m_\rho &= 763  \text{ MeV}
\\
g_{B\sigma}^2 &= 104.3871
\\
g_{B\omega}^2 &= 165.5854
\\
g_{B\rho}^2 &= 79.6
\\
\kappa &= 3.8599 \text{ MeV}
\\
\lambda &= -0.01591.
\end{split}
\end{equation}

In RMF models, all non-fermionic fields are replaced with their mean field values:\ $\sigma \rightarrow \langle \sigma \rangle$, $\omega_\alpha \rightarrow \langle \omega_\alpha \rangle$, and $\vec{\rho}_\alpha \rightarrow \langle \vec{\rho}_\alpha \rangle$ \cite{GlendenningBook}.  This effectively reduces the system to a theory of free fermions which is straightforward to quantize.  We do not review the quantization here, but simply state the results.  The only nonvanishing mean fields are $\langle \sigma \rangle$, $\langle \omega_0 \rangle$, and $\langle \rho_{03} \rangle$.    The energy density and pressure are
\begin{align}
\epsilon &= 
+\frac{1}{2} m_\sigma^2 \langle \sigma \rangle^2 + \frac{\kappa}{3!} (g_{B\sigma} \langle\sigma\rangle)^3 + \frac{\lambda}{4!} (g_{B\sigma} \langle\sigma\rangle)^4
\notag \\
&\qquad + \frac{1}{2} m_\omega^2 \langle\omega_0\rangle^2
+ \frac{1}{2} m_\rho^2 \langle\rho_{03}\rangle^2
\notag \\
&\qquad
+ \frac{1}{8\pi^2}\sum_z
m_{*z}^4
\Bigl[ x_z \sqrt{1+x_z^2}\, ( 2x_z^2 + 1 ) 
\notag \\
&\qquad\qquad\qquad\qquad
 - \ln \left(x_z+\sqrt{1+x_z^2} \right) \Bigr]
\end{align}
and
\begin{align}
p &= - \frac{1}{2} m_\sigma^2 \langle \sigma \rangle^2 - \frac{\kappa}{3!} (g_{B\sigma} \langle\sigma\rangle)^3 - \frac{\lambda}{4!} (g_{B\sigma} \langle\sigma \rangle)^4
\notag \\
&\qquad 
+ \frac{1}{2} m_\omega^2 \langle\omega_0\rangle^2
+ \frac{1}{2} m_\rho^2 \langle\rho_{03}\rangle^2
\notag \\
&\qquad + \frac{1}{24\pi^2}\sum_z  m_{*z}^4 
\Bigl[ x_z \sqrt{1+x_z^2}\, ( 2x_z^2 - 3 ) 
\notag \\
&\qquad\qquad\qquad\qquad
+ 3 \ln \left(x_z+\sqrt{1+x_z^2} \right) \Bigr],
\end{align}
where the sums are over all fields, the general expression for an effective mass is
\begin{equation}
m_{*z} = m_z - g_{z \sigma} \langle\sigma\rangle
\end{equation}
(with $g_{z \sigma} \neq 0$ for $z = p,n$), and $x_z \equiv k_z /m_{*z}$, where $k_z$ is the Fermi momentum for particle $z$.  The number densities are given by $n_z = k_z^3 / 3\pi^2$ and the chemical potentials are
\begin{equation} \label{mu eqs}
\begin{split}
\mu_p &=  g_{B \omega} \langle \omega_0 \rangle + g_{B \rho} I_{3p} \langle\rho_{03}\rangle
+ \sqrt{k_p^2 + m_{*p}^2}
\\
\mu_n &=  g_{B \omega} \langle \omega_0 \rangle + g_{B \rho} I_{3n} \langle\rho_{03}\rangle
+ \sqrt{k_n^2 + m_{*n}^2}
\\
\mu_e &= \sqrt{k_e^2 + m_e^2}
\\
\mu_\mu &= \sqrt{k_\mu^2 + m_{*\mu}^2},
\end{split}
\end{equation}
where $I_{3p} = 1/2$ for the proton and $I_{3n} = -1/2$ for the neutron.  The mean field values are obtained through their equations of motion,
\begin{equation}
\begin{split}
m_\omega^2 \langle \omega_0 \rangle &= g_{B \omega} \sum_{z = p,n}   n_z
\\
m_\rho^2 \langle \rho_{03} \rangle &=g_{B \rho} \sum_{z = p,n}   I_{3z} n_z
\end{split}
\end{equation}
and
\begin{align}
&m_\sigma^2 \langle \sigma \rangle  
+ \frac{\kappa}{2} g_{B \sigma}^3 \langle \sigma\rangle^2 + \frac{\lambda}{6} g_{B \sigma}^4 \langle \sigma\rangle^3
\\
&\quad
= g_{B \sigma} \sum_{z = p,n} 
\frac{m_{*z}^3}{2\pi^2}
\left[ x_z \sqrt{1+x_z^2} - \ln \left(x_z+\sqrt{1+x_z^2} \right) \right].
\notag 
\end{align}

The equation of state, $p(\epsilon)$, for the core is computed by solving the above equations, under the assumption of beta equilibrium, which constrains the chemical potentials to obey $\mu_n = \mu_p + \mu_e$ and $\mu_e = \mu_\mu$, and charge neutrality, which requires $n_p = n_e + n_\mu$, using standard techniques \cite{GlendenningBook}.  At the critical baryonic number density $n_B^c = 0.052$ fm$^{-3}$ \cite{ToddRutel:2005zz}, the core meets the inner crust.  We describe the inner crust using a polytropic equation of state \cite{Carriere:2002bx},
\begin{equation} \label{polytropic eos}
p(\epsilon) = A + B\epsilon^{4/3},
\end{equation}
where $A$ and $B$ are constants.  $A$ and $B$ are determined by matching Eq.\ (\ref{polytropic eos}) with the core at the critical baryonic number density and with the outer crust at the neutron drip line.  For the outer crust we use the Baym\hyp Petchick\hyp Sutherland (BPS) equation of state \cite{Baym:1971pw,Ruester:2005fm}.



\begin{thebibliography}{46}%
\makeatletter
\providecommand \@ifxundefined [1]{%
 \@ifx{#1\undefined}
}%
\providecommand \@ifnum [1]{%
 \ifnum #1\expandafter \@firstoftwo
 \else \expandafter \@secondoftwo
 \fi
}%
\providecommand \@ifx [1]{%
 \ifx #1\expandafter \@firstoftwo
 \else \expandafter \@secondoftwo
 \fi
}%
\providecommand \natexlab [1]{#1}%
\providecommand \enquote  [1]{``#1''}%
\providecommand \bibnamefont  [1]{#1}%
\providecommand \bibfnamefont [1]{#1}%
\providecommand \citenamefont [1]{#1}%
\providecommand \href@noop [0]{\@secondoftwo}%
\providecommand \href [0]{\begingroup \@sanitize@url \@href}%
\providecommand \@href[1]{\@@startlink{#1}\@@href}%
\providecommand \@@href[1]{\endgroup#1\@@endlink}%
\providecommand \@sanitize@url [0]{\catcode `\\12\catcode `\$12\catcode
  `\&12\catcode `\#12\catcode `\^12\catcode `\_12\catcode `\%12\relax}%
\providecommand \@@startlink[1]{}%
\providecommand \@@endlink[0]{}%
\providecommand \url  [0]{\begingroup\@sanitize@url \@url }%
\providecommand \@url [1]{\endgroup\@href {#1}{\urlprefix }}%
\providecommand \urlprefix  [0]{URL }%
\providecommand \Eprint [0]{\href }%
\providecommand \doibase [0]{https://doi.org/}%
\providecommand \selectlanguage [0]{\@gobble}%
\providecommand \bibinfo  [0]{\@secondoftwo}%
\providecommand \bibfield  [0]{\@secondoftwo}%
\providecommand \translation [1]{[#1]}%
\providecommand \BibitemOpen [0]{}%
\providecommand \bibitemStop [0]{}%
\providecommand \bibitemNoStop [0]{.\EOS\space}%
\providecommand \EOS [0]{\spacefactor3000\relax}%
\providecommand \BibitemShut  [1]{\csname bibitem#1\endcsname}%
\let\auto@bib@innerbib\@empty
\bibitem [{\citenamefont {Henriques}\ \emph {et~al.}(1989)\citenamefont
  {Henriques}, \citenamefont {Liddle},\ and\ \citenamefont
  {Moorhouse}}]{Henriques:1989ar}%
  \BibitemOpen
  \bibfield  {author} {\bibinfo {author} {\bibfnamefont {A.~B.}\ \bibnamefont
  {Henriques}}, \bibinfo {author} {\bibfnamefont {A.~R.}\ \bibnamefont
  {Liddle}},\ and\ \bibinfo {author} {\bibfnamefont {R.~G.}\ \bibnamefont
  {Moorhouse}},\ }\bibfield  {title} {\bibinfo {title} {{COMBINED BOSON -
  FERMION STARS}},\ }\href {https://doi.org/10.1016/0370-2693(89)90623-0}
  {\bibfield  {journal} {\bibinfo  {journal} {Phys. Lett. B}\ }\textbf
  {\bibinfo {volume} {233}},\ \bibinfo {pages} {99} (\bibinfo {year}
  {1989})}\BibitemShut {NoStop}%
\bibitem [{\citenamefont {Glendenning}(2000)}]{GlendenningBook}%
  \BibitemOpen
  \bibfield  {author} {\bibinfo {author} {\bibfnamefont {N.~K.}\ \bibnamefont
  {Glendenning}},\ }\href@noop {} {\emph {\bibinfo {title} {{Compact Stars:
  Nuclear Physics, Particle Physics, and General Relativity, 2nd ed.}}}}\
  (\bibinfo  {publisher} {Springer},\ \bibinfo {address} {New York},\ \bibinfo
  {year} {2000})\BibitemShut {NoStop}%
\bibitem [{\citenamefont {Kaup}(1968)}]{Kaup:1968zz}%
  \BibitemOpen
  \bibfield  {author} {\bibinfo {author} {\bibfnamefont {D.~J.}\ \bibnamefont
  {Kaup}},\ }\bibfield  {title} {\bibinfo {title} {{Klein-Gordon Geon}},\
  }\href {https://doi.org/10.1103/PhysRev.172.1331} {\bibfield  {journal}
  {\bibinfo  {journal} {Phys. Rev.}\ }\textbf {\bibinfo {volume} {172}},\
  \bibinfo {pages} {1331} (\bibinfo {year} {1968})}\BibitemShut {NoStop}%
\bibitem [{\citenamefont {Ruffini}\ and\ \citenamefont
  {Bonazzola}(1969)}]{Ruffini:1969qy}%
  \BibitemOpen
  \bibfield  {author} {\bibinfo {author} {\bibfnamefont {R.}~\bibnamefont
  {Ruffini}}\ and\ \bibinfo {author} {\bibfnamefont {S.}~\bibnamefont
  {Bonazzola}},\ }\bibfield  {title} {\bibinfo {title} {{Systems of
  selfgravitating particles in general relativity and the concept of an
  equation of state}},\ }\href {https://doi.org/10.1103/PhysRev.187.1767}
  {\bibfield  {journal} {\bibinfo  {journal} {Phys. Rev.}\ }\textbf {\bibinfo
  {volume} {187}},\ \bibinfo {pages} {1767} (\bibinfo {year}
  {1969})}\BibitemShut {NoStop}%
\bibitem [{\citenamefont {Jetzer}(1992)}]{Jetzer:1991jr}%
  \BibitemOpen
  \bibfield  {author} {\bibinfo {author} {\bibfnamefont {P.}~\bibnamefont
  {Jetzer}},\ }\bibfield  {title} {\bibinfo {title} {{Boson stars}},\ }\href
  {https://doi.org/10.1016/0370-1573(92)90123-H} {\bibfield  {journal}
  {\bibinfo  {journal} {Phys. Rept.}\ }\textbf {\bibinfo {volume} {220}},\
  \bibinfo {pages} {163} (\bibinfo {year} {1992})}\BibitemShut {NoStop}%
\bibitem [{\citenamefont {Schunck}\ and\ \citenamefont
  {Mielke}(2003)}]{Schunck:2003kk}%
  \BibitemOpen
  \bibfield  {author} {\bibinfo {author} {\bibfnamefont {F.~E.}\ \bibnamefont
  {Schunck}}\ and\ \bibinfo {author} {\bibfnamefont {E.~W.}\ \bibnamefont
  {Mielke}},\ }\bibfield  {title} {\bibinfo {title} {{General relativistic
  boson stars}},\ }\href {https://doi.org/10.1088/0264-9381/20/20/201}
  {\bibfield  {journal} {\bibinfo  {journal} {Class. Quant. Grav.}\ }\textbf
  {\bibinfo {volume} {20}},\ \bibinfo {pages} {R301} (\bibinfo {year}
  {2003})},\ \Eprint {https://arxiv.org/abs/0801.0307} {arXiv:0801.0307
  [astro-ph]} \BibitemShut {NoStop}%
\bibitem [{\citenamefont {Liebling}\ and\ \citenamefont
  {Palenzuela}(2012)}]{Liebling:2012fv}%
  \BibitemOpen
  \bibfield  {author} {\bibinfo {author} {\bibfnamefont {S.~L.}\ \bibnamefont
  {Liebling}}\ and\ \bibinfo {author} {\bibfnamefont {C.}~\bibnamefont
  {Palenzuela}},\ }\bibfield  {title} {\bibinfo {title} {{Dynamical Boson
  Stars}},\ }\href {https://doi.org/10.12942/lrr-2012-6} {\bibfield  {journal}
  {\bibinfo  {journal} {Living Rev. Rel.}\ }\textbf {\bibinfo {volume} {15}},\
  \bibinfo {pages} {6} (\bibinfo {year} {2012})},\ \Eprint
  {https://arxiv.org/abs/1202.5809} {arXiv:1202.5809 [gr-qc]} \BibitemShut
  {NoStop}%
\bibitem [{\citenamefont {Akerib}\ \emph {et~al.}(2017)\citenamefont {Akerib}
  \emph {et~al.}}]{Akerib:2017kat}%
  \BibitemOpen
  \bibfield  {author} {\bibinfo {author} {\bibfnamefont {D.}~\bibnamefont
  {Akerib}} \emph {et~al.} (\bibinfo {collaboration} {LUX}),\ }\bibfield
  {title} {\bibinfo {title} {{Limits on spin-dependent WIMP-nucleon cross
  section obtained from the complete LUX exposure}},\ }\href
  {https://doi.org/10.1103/PhysRevLett.118.251302} {\bibfield  {journal}
  {\bibinfo  {journal} {Phys. Rev. Lett.}\ }\textbf {\bibinfo {volume} {118}},\
  \bibinfo {pages} {251302} (\bibinfo {year} {2017})},\ \Eprint
  {https://arxiv.org/abs/1705.03380} {arXiv:1705.03380 [astro-ph.CO]}
  \BibitemShut {NoStop}%
\bibitem [{\citenamefont {Cui}\ \emph {et~al.}(2017)\citenamefont {Cui} \emph
  {et~al.}}]{Cui:2017nnn}%
  \BibitemOpen
  \bibfield  {author} {\bibinfo {author} {\bibfnamefont {X.}~\bibnamefont
  {Cui}} \emph {et~al.} (\bibinfo {collaboration} {PandaX-II}),\ }\bibfield
  {title} {\bibinfo {title} {{Dark Matter Results From 54-Ton-Day Exposure of
  PandaX-II Experiment}},\ }\href
  {https://doi.org/10.1103/PhysRevLett.119.181302} {\bibfield  {journal}
  {\bibinfo  {journal} {Phys. Rev. Lett.}\ }\textbf {\bibinfo {volume} {119}},\
  \bibinfo {pages} {181302} (\bibinfo {year} {2017})},\ \Eprint
  {https://arxiv.org/abs/1708.06917} {arXiv:1708.06917 [astro-ph.CO]}
  \BibitemShut {NoStop}%
\bibitem [{\citenamefont {Aprile}\ \emph {et~al.}(2018)\citenamefont {Aprile}
  \emph {et~al.}}]{Aprile:2018dbl}%
  \BibitemOpen
  \bibfield  {author} {\bibinfo {author} {\bibfnamefont {E.}~\bibnamefont
  {Aprile}} \emph {et~al.} (\bibinfo {collaboration} {XENON}),\ }\bibfield
  {title} {\bibinfo {title} {{Dark Matter Search Results from a One Ton-Year
  Exposure of XENON1T}},\ }\href
  {https://doi.org/10.1103/PhysRevLett.121.111302} {\bibfield  {journal}
  {\bibinfo  {journal} {Phys. Rev. Lett.}\ }\textbf {\bibinfo {volume} {121}},\
  \bibinfo {pages} {111302} (\bibinfo {year} {2018})},\ \Eprint
  {https://arxiv.org/abs/1805.12562} {arXiv:1805.12562 [astro-ph.CO]}
  \BibitemShut {NoStop}%
\bibitem [{\citenamefont {Sandin}\ and\ \citenamefont
  {Ciarcelluti}(2009)}]{Sandin:2008db}%
  \BibitemOpen
  \bibfield  {author} {\bibinfo {author} {\bibfnamefont {F.}~\bibnamefont
  {Sandin}}\ and\ \bibinfo {author} {\bibfnamefont {P.}~\bibnamefont
  {Ciarcelluti}},\ }\bibfield  {title} {\bibinfo {title} {{Effects of mirror
  dark matter on neutron stars}},\ }\href
  {https://doi.org/10.1016/j.astropartphys.2009.09.005} {\bibfield  {journal}
  {\bibinfo  {journal} {Astropart. Phys.}\ }\textbf {\bibinfo {volume} {32}},\
  \bibinfo {pages} {278} (\bibinfo {year} {2009})},\ \Eprint
  {https://arxiv.org/abs/0809.2942} {arXiv:0809.2942 [astro-ph]} \BibitemShut
  {NoStop}%
\bibitem [{\citenamefont {Leung}\ \emph {et~al.}(2011)\citenamefont {Leung},
  \citenamefont {Chu},\ and\ \citenamefont {Lin}}]{Leung:2011zz}%
  \BibitemOpen
  \bibfield  {author} {\bibinfo {author} {\bibfnamefont {S.}~\bibnamefont
  {Leung}}, \bibinfo {author} {\bibfnamefont {M.}~\bibnamefont {Chu}},\ and\
  \bibinfo {author} {\bibfnamefont {L.}~\bibnamefont {Lin}},\ }\bibfield
  {title} {\bibinfo {title} {{Dark-matter admixed neutron stars}},\ }\href
  {https://doi.org/10.1103/PhysRevD.84.107301} {\bibfield  {journal} {\bibinfo
  {journal} {Phys. Rev. D}\ }\textbf {\bibinfo {volume} {84}},\ \bibinfo
  {pages} {107301} (\bibinfo {year} {2011})},\ \Eprint
  {https://arxiv.org/abs/1111.1787} {arXiv:1111.1787 [astro-ph.CO]}
  \BibitemShut {NoStop}%
\bibitem [{\citenamefont {Henriques}\ \emph
  {et~al.}(1990{\natexlab{a}})\citenamefont {Henriques}, \citenamefont
  {Liddle},\ and\ \citenamefont {Moorhouse}}]{Henriques:1989ez}%
  \BibitemOpen
  \bibfield  {author} {\bibinfo {author} {\bibfnamefont {A.~B.}\ \bibnamefont
  {Henriques}}, \bibinfo {author} {\bibfnamefont {A.~R.}\ \bibnamefont
  {Liddle}},\ and\ \bibinfo {author} {\bibfnamefont {R.~G.}\ \bibnamefont
  {Moorhouse}},\ }\bibfield  {title} {\bibinfo {title} {{Combined Boson -
  Fermion Stars: Configurations and Stability}},\ }\href
  {https://doi.org/10.1016/0550-3213(90)90514-E} {\bibfield  {journal}
  {\bibinfo  {journal} {Nucl. Phys. B}\ }\textbf {\bibinfo {volume} {337}},\
  \bibinfo {pages} {737} (\bibinfo {year} {1990}{\natexlab{a}})}\BibitemShut
  {NoStop}%
\bibitem [{\citenamefont {Henriques}\ \emph
  {et~al.}(1990{\natexlab{b}})\citenamefont {Henriques}, \citenamefont
  {Liddle},\ and\ \citenamefont {Moorhouse}}]{Henriques:1990xg}%
  \BibitemOpen
  \bibfield  {author} {\bibinfo {author} {\bibfnamefont {A.~B.}\ \bibnamefont
  {Henriques}}, \bibinfo {author} {\bibfnamefont {A.~R.}\ \bibnamefont
  {Liddle}},\ and\ \bibinfo {author} {\bibfnamefont {R.~G.}\ \bibnamefont
  {Moorhouse}},\ }\bibfield  {title} {\bibinfo {title} {{Stability of boson -
  fermion stars}},\ }\href {https://doi.org/10.1016/0370-2693(90)90789-9}
  {\bibfield  {journal} {\bibinfo  {journal} {Phys. Lett. B}\ }\textbf
  {\bibinfo {volume} {251}},\ \bibinfo {pages} {511} (\bibinfo {year}
  {1990}{\natexlab{b}})}\BibitemShut {NoStop}%
\bibitem [{\citenamefont {Valdez-Alvarado}\ \emph {et~al.}(2013)\citenamefont
  {Valdez-Alvarado}, \citenamefont {Palenzuela}, \citenamefont {Alic},\ and\
  \citenamefont {Ureña-López}}]{ValdezAlvarado:2012xc}%
  \BibitemOpen
  \bibfield  {author} {\bibinfo {author} {\bibfnamefont {S.}~\bibnamefont
  {Valdez-Alvarado}}, \bibinfo {author} {\bibfnamefont {C.}~\bibnamefont
  {Palenzuela}}, \bibinfo {author} {\bibfnamefont {D.}~\bibnamefont {Alic}},\
  and\ \bibinfo {author} {\bibfnamefont {L.~A.}\ \bibnamefont
  {Ureña-López}},\ }\bibfield  {title} {\bibinfo {title} {{Dynamical
  evolution of fermion-boson stars}},\ }\href
  {https://doi.org/10.1103/PhysRevD.87.084040} {\bibfield  {journal} {\bibinfo
  {journal} {Phys. Rev. D}\ }\textbf {\bibinfo {volume} {87}},\ \bibinfo
  {pages} {084040} (\bibinfo {year} {2013})},\ \Eprint
  {https://arxiv.org/abs/1210.2299} {arXiv:1210.2299 [gr-qc]} \BibitemShut
  {NoStop}%
\bibitem [{\citenamefont {Di~Giovanni}\ \emph {et~al.}(2020)\citenamefont
  {Di~Giovanni}, \citenamefont {Fakhry}, \citenamefont {Sanchis-Gual},
  \citenamefont {Degollado},\ and\ \citenamefont {Font}}]{DiGiovanni:2020frc}%
  \BibitemOpen
  \bibfield  {author} {\bibinfo {author} {\bibfnamefont {F.}~\bibnamefont
  {Di~Giovanni}}, \bibinfo {author} {\bibfnamefont {S.}~\bibnamefont {Fakhry}},
  \bibinfo {author} {\bibfnamefont {N.}~\bibnamefont {Sanchis-Gual}}, \bibinfo
  {author} {\bibfnamefont {J.~C.}\ \bibnamefont {Degollado}},\ and\ \bibinfo
  {author} {\bibfnamefont {J.~A.}\ \bibnamefont {Font}},\ }\bibfield  {title}
  {\bibinfo {title} {{Dynamical formation and stability of fermion-boson
  stars}},\ }\href {https://doi.org/10.1103/PhysRevD.102.084063} {\bibfield
  {journal} {\bibinfo  {journal} {Phys. Rev. D}\ }\textbf {\bibinfo {volume}
  {102}},\ \bibinfo {pages} {084063} (\bibinfo {year} {2020})},\ \Eprint
  {https://arxiv.org/abs/2006.08583} {arXiv:2006.08583 [gr-qc]} \BibitemShut
  {NoStop}%
\bibitem [{\citenamefont {Valdez-Alvarado}\ \emph {et~al.}(2020)\citenamefont
  {Valdez-Alvarado}, \citenamefont {Becerril},\ and\ \citenamefont {Ure\~na
  L\'opez}}]{Valdez-Alvarado:2020vqa}%
  \BibitemOpen
  \bibfield  {author} {\bibinfo {author} {\bibfnamefont {S.}~\bibnamefont
  {Valdez-Alvarado}}, \bibinfo {author} {\bibfnamefont {R.}~\bibnamefont
  {Becerril}},\ and\ \bibinfo {author} {\bibfnamefont {L.~A.}\ \bibnamefont
  {Ure\~na L\'opez}},\ }\bibfield  {title} {\bibinfo {title} {{Fermion-boson
  stars with a quartic self-interaction in the boson sector}},\ }\href
  {https://doi.org/10.1103/PhysRevD.102.064038} {\bibfield  {journal} {\bibinfo
   {journal} {Phys. Rev. D}\ }\textbf {\bibinfo {volume} {102}},\ \bibinfo
  {pages} {064038} (\bibinfo {year} {2020})},\ \Eprint
  {https://arxiv.org/abs/2001.11009} {arXiv:2001.11009 [gr-qc]} \BibitemShut
  {NoStop}%
\bibitem [{\citenamefont {Di~Giovanni}\ \emph {et~al.}(2021)\citenamefont
  {Di~Giovanni}, \citenamefont {Fakhry}, \citenamefont {Sanchis-Gual},
  \citenamefont {Degollado},\ and\ \citenamefont {Font}}]{DiGiovanni:2021vlu}%
  \BibitemOpen
  \bibfield  {author} {\bibinfo {author} {\bibfnamefont {F.}~\bibnamefont
  {Di~Giovanni}}, \bibinfo {author} {\bibfnamefont {S.}~\bibnamefont {Fakhry}},
  \bibinfo {author} {\bibfnamefont {N.}~\bibnamefont {Sanchis-Gual}}, \bibinfo
  {author} {\bibfnamefont {J.~C.}\ \bibnamefont {Degollado}},\ and\ \bibinfo
  {author} {\bibfnamefont {J.~A.}\ \bibnamefont {Font}},\ }\bibfield  {title}
  {\bibinfo {title} {{A stabilization mechanism for excited fermion-boson
  stars}},\ }\href@noop {} {\  (\bibinfo {year} {2021})},\ \Eprint
  {https://arxiv.org/abs/2105.00530} {arXiv:2105.00530 [gr-qc]} \BibitemShut
  {NoStop}%
\bibitem [{\citenamefont {de~Sousa}\ and\ \citenamefont
  {Tomazelli}(1998)}]{deSousa:1995ye}%
  \BibitemOpen
  \bibfield  {author} {\bibinfo {author} {\bibfnamefont {C.~M.~G.}\
  \bibnamefont {de~Sousa}}\ and\ \bibinfo {author} {\bibfnamefont {J.~L.}\
  \bibnamefont {Tomazelli}},\ }\bibfield  {title} {\bibinfo {title} {{A Model
  for stars of interacting bosons and fermions}},\ }\href
  {https://doi.org/10.1103/PhysRevD.58.123003} {\bibfield  {journal} {\bibinfo
  {journal} {Phys. Rev. D}\ }\textbf {\bibinfo {volume} {58}},\ \bibinfo
  {pages} {123003} (\bibinfo {year} {1998})},\ \Eprint
  {https://arxiv.org/abs/gr-qc/9507043} {arXiv:gr-qc/9507043} \BibitemShut
  {NoStop}%
\bibitem [{\citenamefont {Pisano}\ and\ \citenamefont
  {Tomazelli}(1996)}]{Pisano:1995yk}%
  \BibitemOpen
  \bibfield  {author} {\bibinfo {author} {\bibfnamefont {F.}~\bibnamefont
  {Pisano}}\ and\ \bibinfo {author} {\bibfnamefont {J.~L.}\ \bibnamefont
  {Tomazelli}},\ }\bibfield  {title} {\bibinfo {title} {{Stars of WIMPs}},\
  }\href {https://doi.org/10.1142/S0217732396000667} {\bibfield  {journal}
  {\bibinfo  {journal} {Mod. Phys. Lett. A}\ }\textbf {\bibinfo {volume}
  {11}},\ \bibinfo {pages} {647} (\bibinfo {year} {1996})},\ \Eprint
  {https://arxiv.org/abs/gr-qc/9509022} {arXiv:gr-qc/9509022} \BibitemShut
  {NoStop}%
\bibitem [{\citenamefont {Sakamoto}\ and\ \citenamefont
  {Shiraishi}(1998)}]{Sakamoto:1998aj}%
  \BibitemOpen
  \bibfield  {author} {\bibinfo {author} {\bibfnamefont {K.}~\bibnamefont
  {Sakamoto}}\ and\ \bibinfo {author} {\bibfnamefont {K.}~\bibnamefont
  {Shiraishi}},\ }\bibfield  {title} {\bibinfo {title} {{Exact solutions for
  boson fermion stars in (2+1)-dimensions}},\ }\href
  {https://doi.org/10.1103/PhysRevD.58.124017} {\bibfield  {journal} {\bibinfo
  {journal} {Phys. Rev. D}\ }\textbf {\bibinfo {volume} {58}},\ \bibinfo
  {pages} {124017} (\bibinfo {year} {1998})},\ \Eprint
  {https://arxiv.org/abs/gr-qc/9806040} {arXiv:gr-qc/9806040} \BibitemShut
  {NoStop}%
\bibitem [{\citenamefont {de~Sousa}\ and\ \citenamefont
  {Silveira}(2001)}]{deSousa:2000eq}%
  \BibitemOpen
  \bibfield  {author} {\bibinfo {author} {\bibfnamefont {C.~M.~G.}\
  \bibnamefont {de~Sousa}}\ and\ \bibinfo {author} {\bibfnamefont
  {V.}~\bibnamefont {Silveira}},\ }\bibfield  {title} {\bibinfo {title}
  {{Slowly rotating boson fermion stars}},\ }\href
  {https://doi.org/10.1142/S0218271801001360} {\bibfield  {journal} {\bibinfo
  {journal} {Int. J. Mod. Phys. D}\ }\textbf {\bibinfo {volume} {10}},\
  \bibinfo {pages} {881} (\bibinfo {year} {2001})},\ \Eprint
  {https://arxiv.org/abs/gr-qc/0012020} {arXiv:gr-qc/0012020} \BibitemShut
  {NoStop}%
\bibitem [{\citenamefont {Henriques}\ and\ \citenamefont
  {Mendes}(2005)}]{Henriques:2003yr}%
  \BibitemOpen
  \bibfield  {author} {\bibinfo {author} {\bibfnamefont {A.~B.}\ \bibnamefont
  {Henriques}}\ and\ \bibinfo {author} {\bibfnamefont {L.~E.}\ \bibnamefont
  {Mendes}},\ }\bibfield  {title} {\bibinfo {title} {{Boson - fermion stars:
  Exploring different configurations}},\ }\href
  {https://doi.org/10.1007/s10509-005-4512-1} {\bibfield  {journal} {\bibinfo
  {journal} {Astrophys. Space Sci.}\ }\textbf {\bibinfo {volume} {300}},\
  \bibinfo {pages} {367} (\bibinfo {year} {2005})},\ \Eprint
  {https://arxiv.org/abs/astro-ph/0301015} {arXiv:astro-ph/0301015}
  \BibitemShut {NoStop}%
\bibitem [{\citenamefont {Dzhunushaliev}\ \emph {et~al.}(2011)\citenamefont
  {Dzhunushaliev}, \citenamefont {Folomeev},\ and\ \citenamefont
  {Singleton}}]{Dzhunushaliev:2011ma}%
  \BibitemOpen
  \bibfield  {author} {\bibinfo {author} {\bibfnamefont {V.}~\bibnamefont
  {Dzhunushaliev}}, \bibinfo {author} {\bibfnamefont {V.}~\bibnamefont
  {Folomeev}},\ and\ \bibinfo {author} {\bibfnamefont {D.}~\bibnamefont
  {Singleton}},\ }\bibfield  {title} {\bibinfo {title} {{Chameleon stars}},\
  }\href {https://doi.org/10.1103/PhysRevD.84.084025} {\bibfield  {journal}
  {\bibinfo  {journal} {Phys. Rev. D}\ }\textbf {\bibinfo {volume} {84}},\
  \bibinfo {pages} {084025} (\bibinfo {year} {2011})},\ \Eprint
  {https://arxiv.org/abs/1106.1267} {arXiv:1106.1267 [astro-ph.SR]}
  \BibitemShut {NoStop}%
\bibitem [{\citenamefont {Brito}\ \emph {et~al.}(2015)\citenamefont {Brito},
  \citenamefont {Cardoso},\ and\ \citenamefont {Okawa}}]{Brito:2015yga}%
  \BibitemOpen
  \bibfield  {author} {\bibinfo {author} {\bibfnamefont {R.}~\bibnamefont
  {Brito}}, \bibinfo {author} {\bibfnamefont {V.}~\bibnamefont {Cardoso}},\
  and\ \bibinfo {author} {\bibfnamefont {H.}~\bibnamefont {Okawa}},\ }\bibfield
   {title} {\bibinfo {title} {{Accretion of dark matter by stars}},\ }\href
  {https://doi.org/10.1103/PhysRevLett.115.111301} {\bibfield  {journal}
  {\bibinfo  {journal} {Phys. Rev. Lett.}\ }\textbf {\bibinfo {volume} {115}},\
  \bibinfo {pages} {111301} (\bibinfo {year} {2015})},\ \Eprint
  {https://arxiv.org/abs/1508.04773} {arXiv:1508.04773 [gr-qc]} \BibitemShut
  {NoStop}%
\bibitem [{\citenamefont {Brito}\ \emph {et~al.}(2016)\citenamefont {Brito},
  \citenamefont {Cardoso}, \citenamefont {Macedo}, \citenamefont {Okawa},\ and\
  \citenamefont {Palenzuela}}]{Brito:2015yfh}%
  \BibitemOpen
  \bibfield  {author} {\bibinfo {author} {\bibfnamefont {R.}~\bibnamefont
  {Brito}}, \bibinfo {author} {\bibfnamefont {V.}~\bibnamefont {Cardoso}},
  \bibinfo {author} {\bibfnamefont {C.~F.~B.}\ \bibnamefont {Macedo}}, \bibinfo
  {author} {\bibfnamefont {H.}~\bibnamefont {Okawa}},\ and\ \bibinfo {author}
  {\bibfnamefont {C.}~\bibnamefont {Palenzuela}},\ }\bibfield  {title}
  {\bibinfo {title} {{Interaction between bosonic dark matter and stars}},\
  }\href {https://doi.org/10.1103/PhysRevD.93.044045} {\bibfield  {journal}
  {\bibinfo  {journal} {Phys. Rev. D}\ }\textbf {\bibinfo {volume} {93}},\
  \bibinfo {pages} {044045} (\bibinfo {year} {2016})},\ \Eprint
  {https://arxiv.org/abs/1512.00466} {arXiv:1512.00466 [astro-ph.SR]}
  \BibitemShut {NoStop}%
\bibitem [{\citenamefont {Jetzer}\ and\ \citenamefont {van~der
  Bij}(1989)}]{Jetzer:1989av}%
  \BibitemOpen
  \bibfield  {author} {\bibinfo {author} {\bibfnamefont {P.}~\bibnamefont
  {Jetzer}}\ and\ \bibinfo {author} {\bibfnamefont {J.~J.}\ \bibnamefont
  {van~der Bij}},\ }\bibfield  {title} {\bibinfo {title} {{CHARGED BOSON
  STARS}},\ }\href {https://doi.org/10.1016/0370-2693(89)90941-6} {\bibfield
  {journal} {\bibinfo  {journal} {Phys. Lett. B}\ }\textbf {\bibinfo {volume}
  {227}},\ \bibinfo {pages} {341} (\bibinfo {year} {1989})}\BibitemShut
  {NoStop}%
\bibitem [{\citenamefont {Pugliese}\ \emph {et~al.}(2013)\citenamefont
  {Pugliese}, \citenamefont {Quevedo}, \citenamefont {Rueda~H.},\ and\
  \citenamefont {Ruffini}}]{Pugliese:2013gsa}%
  \BibitemOpen
  \bibfield  {author} {\bibinfo {author} {\bibfnamefont {D.}~\bibnamefont
  {Pugliese}}, \bibinfo {author} {\bibfnamefont {H.}~\bibnamefont {Quevedo}},
  \bibinfo {author} {\bibfnamefont {J.~A.}\ \bibnamefont {Rueda~H.}},\ and\
  \bibinfo {author} {\bibfnamefont {R.}~\bibnamefont {Ruffini}},\ }\bibfield
  {title} {\bibinfo {title} {{On charged boson stars}},\ }\href
  {https://doi.org/10.1103/PhysRevD.88.024053} {\bibfield  {journal} {\bibinfo
  {journal} {Phys. Rev. D}\ }\textbf {\bibinfo {volume} {88}},\ \bibinfo
  {pages} {024053} (\bibinfo {year} {2013})},\ \Eprint
  {https://arxiv.org/abs/1305.4241} {arXiv:1305.4241 [astro-ph.HE]}
  \BibitemShut {NoStop}%
\bibitem [{\citenamefont {Sin}(1994)}]{Sin:1992bg}%
  \BibitemOpen
  \bibfield  {author} {\bibinfo {author} {\bibfnamefont {S.-J.}\ \bibnamefont
  {Sin}},\ }\bibfield  {title} {\bibinfo {title} {{Late time cosmological phase
  transition and galactic halo as Bose liquid}},\ }\href
  {https://doi.org/10.1103/PhysRevD.50.3650} {\bibfield  {journal} {\bibinfo
  {journal} {Phys. Rev. D}\ }\textbf {\bibinfo {volume} {50}},\ \bibinfo
  {pages} {3650} (\bibinfo {year} {1994})},\ \Eprint
  {https://arxiv.org/abs/hep-ph/9205208} {arXiv:hep-ph/9205208} \BibitemShut
  {NoStop}%
\bibitem [{\citenamefont {Chavanis}\ and\ \citenamefont
  {Harko}(2012)}]{Chavanis:2011cz}%
  \BibitemOpen
  \bibfield  {author} {\bibinfo {author} {\bibfnamefont {P.-H.}\ \bibnamefont
  {Chavanis}}\ and\ \bibinfo {author} {\bibfnamefont {T.}~\bibnamefont
  {Harko}},\ }\bibfield  {title} {\bibinfo {title} {{Bose-Einstein Condensate
  general relativistic stars}},\ }\href
  {https://doi.org/10.1103/PhysRevD.86.064011} {\bibfield  {journal} {\bibinfo
  {journal} {Phys. Rev. D}\ }\textbf {\bibinfo {volume} {86}},\ \bibinfo
  {pages} {064011} (\bibinfo {year} {2012})},\ \Eprint
  {https://arxiv.org/abs/1108.3986} {arXiv:1108.3986 [astro-ph.SR]}
  \BibitemShut {NoStop}%
\bibitem [{\citenamefont {Hu}\ \emph {et~al.}(2000)\citenamefont {Hu},
  \citenamefont {Barkana},\ and\ \citenamefont {Gruzinov}}]{Hu:2000ke}%
  \BibitemOpen
  \bibfield  {author} {\bibinfo {author} {\bibfnamefont {W.}~\bibnamefont
  {Hu}}, \bibinfo {author} {\bibfnamefont {R.}~\bibnamefont {Barkana}},\ and\
  \bibinfo {author} {\bibfnamefont {A.}~\bibnamefont {Gruzinov}},\ }\bibfield
  {title} {\bibinfo {title} {{Cold and fuzzy dark matter}},\ }\href
  {https://doi.org/10.1103/PhysRevLett.85.1158} {\bibfield  {journal} {\bibinfo
   {journal} {Phys. Rev. Lett.}\ }\textbf {\bibinfo {volume} {85}},\ \bibinfo
  {pages} {1158} (\bibinfo {year} {2000})},\ \Eprint
  {https://arxiv.org/abs/astro-ph/0003365} {arXiv:astro-ph/0003365}
  \BibitemShut {NoStop}%
\bibitem [{\citenamefont {Lalazissis}\ \emph {et~al.}(1997)\citenamefont
  {Lalazissis}, \citenamefont {Konig},\ and\ \citenamefont
  {Ring}}]{Lalazissis:1996rd}%
  \BibitemOpen
  \bibfield  {author} {\bibinfo {author} {\bibfnamefont {G.~A.}\ \bibnamefont
  {Lalazissis}}, \bibinfo {author} {\bibfnamefont {J.}~\bibnamefont {Konig}},\
  and\ \bibinfo {author} {\bibfnamefont {P.}~\bibnamefont {Ring}},\ }\bibfield
  {title} {\bibinfo {title} {{A New parametrization for the Lagrangian density
  of relativistic mean field theory}},\ }\href
  {https://doi.org/10.1103/PhysRevC.55.540} {\bibfield  {journal} {\bibinfo
  {journal} {Phys. Rev. C}\ }\textbf {\bibinfo {volume} {55}},\ \bibinfo
  {pages} {540} (\bibinfo {year} {1997})},\ \Eprint
  {https://arxiv.org/abs/nucl-th/9607039} {arXiv:nucl-th/9607039} \BibitemShut
  {NoStop}%
\bibitem [{\citenamefont {Todd-Rutel}\ and\ \citenamefont
  {Piekarewicz}(2005)}]{ToddRutel:2005zz}%
  \BibitemOpen
  \bibfield  {author} {\bibinfo {author} {\bibfnamefont {B.~G.}\ \bibnamefont
  {Todd-Rutel}}\ and\ \bibinfo {author} {\bibfnamefont {J.}~\bibnamefont
  {Piekarewicz}},\ }\bibfield  {title} {\bibinfo {title} {{Neutron-Rich Nuclei
  and Neutron Stars: A New Accurately Calibrated Interaction for the Study of
  Neutron-Rich Matter}},\ }\href
  {https://doi.org/10.1103/PhysRevLett.95.122501} {\bibfield  {journal}
  {\bibinfo  {journal} {Phys. Rev. Lett.}\ }\textbf {\bibinfo {volume} {95}},\
  \bibinfo {pages} {122501} (\bibinfo {year} {2005})},\ \Eprint
  {https://arxiv.org/abs/nucl-th/0504034} {arXiv:nucl-th/0504034} \BibitemShut
  {NoStop}%
\bibitem [{\citenamefont {Chandrasekhar}(1964)}]{Chandrasekhar:1964zz}%
  \BibitemOpen
  \bibfield  {author} {\bibinfo {author} {\bibfnamefont {S.}~\bibnamefont
  {Chandrasekhar}},\ }\bibfield  {title} {\bibinfo {title} {{The Dynamical
  Instability of Gaseous Masses Approaching the Schwarzschild Limit in General
  Relativity}},\ }\href {https://doi.org/10.1086/147938} {\bibfield  {journal}
  {\bibinfo  {journal} {Astrophys. J.}\ }\textbf {\bibinfo {volume} {140}},\
  \bibinfo {pages} {417} (\bibinfo {year} {1964})},\ \bibinfo {note} {[Erratum:
  Astrophys.J. 140, 1342 (1964)]}\BibitemShut {NoStop}%
\bibitem [{\citenamefont {Gleiser}(1988)}]{Gleiser:1988rq}%
  \BibitemOpen
  \bibfield  {author} {\bibinfo {author} {\bibfnamefont {M.}~\bibnamefont
  {Gleiser}},\ }\bibfield  {title} {\bibinfo {title} {{Stability of Boson
  Stars}},\ }\href {https://doi.org/10.1103/PhysRevD.38.2376} {\bibfield
  {journal} {\bibinfo  {journal} {Phys. Rev. D}\ }\textbf {\bibinfo {volume}
  {38}},\ \bibinfo {pages} {2376} (\bibinfo {year} {1988})},\ \bibinfo {note}
  {[Erratum: Phys.Rev.D 39, 1257 (1989)]}\BibitemShut {NoStop}%
\bibitem [{\citenamefont {Jetzer}(1989{\natexlab{a}})}]{Jetzer:1988vr}%
  \BibitemOpen
  \bibfield  {author} {\bibinfo {author} {\bibfnamefont {P.}~\bibnamefont
  {Jetzer}},\ }\bibfield  {title} {\bibinfo {title} {{Dynamical Instability of
  Bosonic Stellar Configurations}},\ }\href
  {https://doi.org/10.1016/0550-3213(89)90038-2} {\bibfield  {journal}
  {\bibinfo  {journal} {Nucl. Phys. B}\ }\textbf {\bibinfo {volume} {316}},\
  \bibinfo {pages} {411} (\bibinfo {year} {1989}{\natexlab{a}})}\BibitemShut
  {NoStop}%
\bibitem [{\citenamefont {Gleiser}\ and\ \citenamefont
  {Watkins}(1989)}]{Gleiser:1988ih}%
  \BibitemOpen
  \bibfield  {author} {\bibinfo {author} {\bibfnamefont {M.}~\bibnamefont
  {Gleiser}}\ and\ \bibinfo {author} {\bibfnamefont {R.}~\bibnamefont
  {Watkins}},\ }\bibfield  {title} {\bibinfo {title} {{Gravitational Stability
  of Scalar Matter}},\ }\href {https://doi.org/10.1016/0550-3213(89)90627-5}
  {\bibfield  {journal} {\bibinfo  {journal} {Nucl. Phys. B}\ }\textbf
  {\bibinfo {volume} {319}},\ \bibinfo {pages} {733} (\bibinfo {year}
  {1989})}\BibitemShut {NoStop}%
\bibitem [{\citenamefont {Lee}\ and\ \citenamefont {Pang}(1989)}]{Lee:1988av}%
  \BibitemOpen
  \bibfield  {author} {\bibinfo {author} {\bibfnamefont {T.~D.}\ \bibnamefont
  {Lee}}\ and\ \bibinfo {author} {\bibfnamefont {Y.}~\bibnamefont {Pang}},\
  }\bibfield  {title} {\bibinfo {title} {{Stability of Mini - Boson Stars}},\
  }\href {https://doi.org/10.1016/0550-3213(89)90365-9} {\bibfield  {journal}
  {\bibinfo  {journal} {Nucl. Phys. B}\ }\textbf {\bibinfo {volume} {315}},\
  \bibinfo {pages} {477} (\bibinfo {year} {1989})},\ \bibinfo {note}
  {[,129(1988)]}\BibitemShut {NoStop}%
\bibitem [{\citenamefont {Hawley}\ and\ \citenamefont
  {Choptuik}(2000)}]{Hawley:2000dt}%
  \BibitemOpen
  \bibfield  {author} {\bibinfo {author} {\bibfnamefont {S.~H.}\ \bibnamefont
  {Hawley}}\ and\ \bibinfo {author} {\bibfnamefont {M.~W.}\ \bibnamefont
  {Choptuik}},\ }\bibfield  {title} {\bibinfo {title} {{Boson stars driven to
  the brink of black hole formation}},\ }\href
  {https://doi.org/10.1103/PhysRevD.62.104024} {\bibfield  {journal} {\bibinfo
  {journal} {Phys. Rev. D}\ }\textbf {\bibinfo {volume} {62}},\ \bibinfo
  {pages} {104024} (\bibinfo {year} {2000})},\ \Eprint
  {https://arxiv.org/abs/gr-qc/0007039} {arXiv:gr-qc/0007039} \BibitemShut
  {NoStop}%
\bibitem [{\citenamefont {Kain}(2021{\natexlab{a}})}]{kain}%
  \BibitemOpen
  \bibfield  {author} {\bibinfo {author} {\bibfnamefont {B.}~\bibnamefont
  {Kain}},\ }\bibfield  {title} {\bibinfo {title} {{Boson stars and their
  radial oscillations}},\ }\href {https://doi.org/10.1103/PhysRevD.103.123003}
  {\bibfield  {journal} {\bibinfo  {journal} {Phys. Rev. D}\ }\textbf {\bibinfo
  {volume} {103}},\ \bibinfo {pages} {123003} (\bibinfo {year}
  {2021}{\natexlab{a}})},\ \Eprint {https://arxiv.org/abs/2106.01740}
  {arXiv:2106.01740 [gr-qc]} \BibitemShut {NoStop}%
\bibitem [{\citenamefont {Jetzer}(1989{\natexlab{b}})}]{Jetzer:1989us}%
  \BibitemOpen
  \bibfield  {author} {\bibinfo {author} {\bibfnamefont {P.}~\bibnamefont
  {Jetzer}},\ }\bibfield  {title} {\bibinfo {title} {{Stability of Charged
  Boson Stars}},\ }\href {https://doi.org/10.1016/0370-2693(89)90689-8}
  {\bibfield  {journal} {\bibinfo  {journal} {Phys. Lett. B}\ }\textbf
  {\bibinfo {volume} {231}},\ \bibinfo {pages} {433} (\bibinfo {year}
  {1989}{\natexlab{b}})}\BibitemShut {NoStop}%
\bibitem [{\citenamefont {Kain}(2021{\natexlab{b}})}]{Kain:2021hpk}%
  \BibitemOpen
  \bibfield  {author} {\bibinfo {author} {\bibfnamefont {B.}~\bibnamefont
  {Kain}},\ }\bibfield  {title} {\bibinfo {title} {{Dark matter admixed neutron
  stars}},\ }\href {https://doi.org/10.1103/PhysRevD.103.043009} {\bibfield
  {journal} {\bibinfo  {journal} {Phys. Rev. D}\ }\textbf {\bibinfo {volume}
  {103}},\ \bibinfo {pages} {043009} (\bibinfo {year} {2021}{\natexlab{b}})},\
  \Eprint {https://arxiv.org/abs/2102.08257} {arXiv:2102.08257 [gr-qc]}
  \BibitemShut {NoStop}%
\bibitem [{\citenamefont {Jetzer}(1990)}]{Jetzer:1990xa}%
  \BibitemOpen
  \bibfield  {author} {\bibinfo {author} {\bibfnamefont {P.}~\bibnamefont
  {Jetzer}},\ }\bibfield  {title} {\bibinfo {title} {{Stability of Combined
  Boson - Fermion Stars}},\ }\href
  {https://doi.org/10.1016/0370-2693(90)90952-3} {\bibfield  {journal}
  {\bibinfo  {journal} {Phys. Lett. B}\ }\textbf {\bibinfo {volume} {243}},\
  \bibinfo {pages} {36} (\bibinfo {year} {1990})}\BibitemShut {NoStop}%
\bibitem [{\citenamefont {Carriere}\ \emph {et~al.}(2003)\citenamefont
  {Carriere}, \citenamefont {Horowitz},\ and\ \citenamefont
  {Piekarewicz}}]{Carriere:2002bx}%
  \BibitemOpen
  \bibfield  {author} {\bibinfo {author} {\bibfnamefont {J.}~\bibnamefont
  {Carriere}}, \bibinfo {author} {\bibfnamefont {C.~J.}\ \bibnamefont
  {Horowitz}},\ and\ \bibinfo {author} {\bibfnamefont {J.}~\bibnamefont
  {Piekarewicz}},\ }\bibfield  {title} {\bibinfo {title} {{Low mass neutron
  stars and the equation of state of dense matter}},\ }\href
  {https://doi.org/10.1086/376515} {\bibfield  {journal} {\bibinfo  {journal}
  {Astrophys. J.}\ }\textbf {\bibinfo {volume} {593}},\ \bibinfo {pages} {463}
  (\bibinfo {year} {2003})},\ \Eprint {https://arxiv.org/abs/nucl-th/0211015}
  {arXiv:nucl-th/0211015} \BibitemShut {NoStop}%
\bibitem [{\citenamefont {Baym}\ \emph {et~al.}(1971)\citenamefont {Baym},
  \citenamefont {Pethick},\ and\ \citenamefont {Sutherland}}]{Baym:1971pw}%
  \BibitemOpen
  \bibfield  {author} {\bibinfo {author} {\bibfnamefont {G.}~\bibnamefont
  {Baym}}, \bibinfo {author} {\bibfnamefont {C.}~\bibnamefont {Pethick}},\ and\
  \bibinfo {author} {\bibfnamefont {P.}~\bibnamefont {Sutherland}},\ }\bibfield
   {title} {\bibinfo {title} {{The Ground state of matter at high densities:
  Equation of state and stellar models}},\ }\href
  {https://doi.org/10.1086/151216} {\bibfield  {journal} {\bibinfo  {journal}
  {Astrophys. J.}\ }\textbf {\bibinfo {volume} {170}},\ \bibinfo {pages} {299}
  (\bibinfo {year} {1971})}\BibitemShut {NoStop}%
\bibitem [{\citenamefont {Ruester}\ \emph {et~al.}(2006)\citenamefont
  {Ruester}, \citenamefont {Hempel},\ and\ \citenamefont
  {Schaffner-Bielich}}]{Ruester:2005fm}%
  \BibitemOpen
  \bibfield  {author} {\bibinfo {author} {\bibfnamefont {S.~B.}\ \bibnamefont
  {Ruester}}, \bibinfo {author} {\bibfnamefont {M.}~\bibnamefont {Hempel}},\
  and\ \bibinfo {author} {\bibfnamefont {J.}~\bibnamefont
  {Schaffner-Bielich}},\ }\bibfield  {title} {\bibinfo {title} {{The outer
  crust of non-accreting cold neutron stars}},\ }\href
  {https://doi.org/10.1103/PhysRevC.73.035804} {\bibfield  {journal} {\bibinfo
  {journal} {Phys. Rev. C}\ }\textbf {\bibinfo {volume} {73}},\ \bibinfo
  {pages} {035804} (\bibinfo {year} {2006})},\ \Eprint
  {https://arxiv.org/abs/astro-ph/0509325} {arXiv:astro-ph/0509325}
  \BibitemShut {NoStop}%
\end{thebibliography}

%

\end{document}